\documentclass[lettersize,journal]{amsart}
\usepackage{dsfont}
\usepackage[utf8]{inputenc}
\usepackage[compress]{cite}
\usepackage{graphicx}
\usepackage{amsmath}
\usepackage{amssymb}
\usepackage{mathtools}
\usepackage{algorithmic}
\usepackage{subfigure} 
\usepackage{xr}
\usepackage{datatool}
\usepackage{steinmetz}
\usepackage{setspace}
\usepackage{xcolor}
\usepackage{array}% http://ctan.org/pkg/array
\usepackage{tabularx}
\usepackage{comment}
\usepackage{hyperref}
\usepackage{booktabs}
\usepackage{multirow}
\usepackage{flushend}
\usepackage{titleps}
\usepackage{adjustbox}

\newcommand{\mae}{\operatorname{MAE}}

\begin{document}
\title[Enhancing Missing Data Imputation]{Enhancing Missing Data Imputation of Non-stationary Signals with Harmonic Decomposition}
	
\author[J Ruiz, HT Wu, and MA Colominas]{Joaquin Ruiz, Hau-Tieng Wu, and Marcelo A. Colominas}

\begin{abstract}
Dealing with time series with missing values, including those afflicted by low quality or over-saturation, presents a significant signal processing challenge. The task of recovering these missing values, known as imputation, has led to the development of several algorithms. However, we have observed that the efficacy of these algorithms tends to diminish when the time series exhibit non-stationary oscillatory behavior.
In this paper, we introduce a novel algorithm, coined Harmonic Level Interpolation (\textsf{HaLI}), which enhances the performance of existing imputation algorithms for oscillatory time series.
After running any chosen imputation algorithm, \textsf{HaLI} leverages the {\em harmonic decomposition} based on the {\em adaptive nonharmonic model} of the initial imputation to improve the imputation accuracy for oscillatory time series.
Experimental assessments conducted on synthetic and real signals consistently highlight that \textsf{HaLI} enhances the performance of existing imputation algorithms.
The algorithm is made publicly available as a readily employable Matlab code for other researchers to use.
\end{abstract}

\maketitle

\section{Introduction}

Missing data is a pervasive issue \cite{jelivcic2009use,enders2022applied} that necessitates different strategies for handling depending on the dataset type and data analysis objectives, presenting various challenges to researchers. This paper centers on the predicament of missing values within {\em oscillatory time series} and the exploration of imputation methods for their retrieval.  This issue frequently surfaces in digital health, particularly during extended health monitoring using biomedical sensors. Factors such as patient motion, sensor disruptions, and calibration hurdles can lead to absent values in recorded biomedical time series. While akin challenges exist in other fields, our focus predominantly rests on biomedical applications. Notably, the insights we provide can be extended to relevant time series analysis in different domains.

\subsection{Categorization of missing value}

A traditional classification of missing data was proposed in \cite{rubin1976inference,ibrahim2012missing} for the general missing value problem that treats the missing data indicators as random variables. The description depends on the observed and unobserved values in a dataset.  {\em Missing completely at random} (MCAR) refers to cases where the data missingness does not depend on the data, either observed or unobserved. Note that it does not mean that the missing pattern is random. {\em Missing at random} (MAR) refers to cases where the data missingness does not depend on unobserved data. MAR along with the assumption that the parameters of the missing data mechanism and sampling are distinct is called {\em ignorable missing}. Finally, {\em missing not at random} (MNAR) means that the data missingness depends on the observed data, or if it is non-ignorable. See \cite{rubin1976inference,little2019statistical} for a rigorous mathematical definition and \cite{ibrahim2012missing} for concrete examples in clinical studies.

\begin{figure}[t!]
	\centering
	\includegraphics[trim=0 20 0 0, width=\columnwidth]{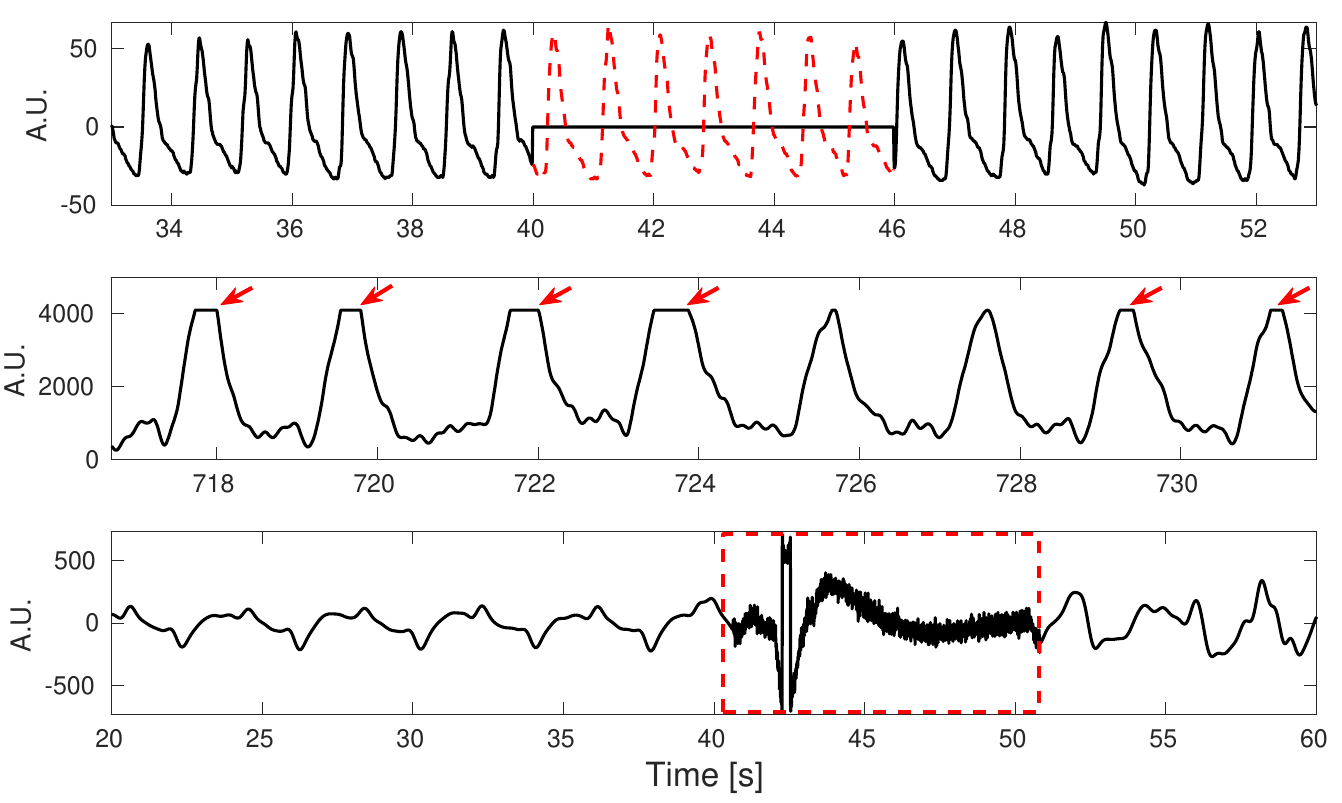}
	\caption{Examples of missing data in the biomedical field. Top: Photoplethysmography signal depicting missing data resulting from sensor disconnection. Middle: Oversaturated airflow signal featuring saturated peaks highlighted by red arrows. Bottom: Airflow signal showcasing a low-quality segment emphasized by the red box.}
	\label{fig:miss_types}
\end{figure}

The above classification is generic and does not consider the characteristics of time series. While this classification becomes feasible when a suitable statistical model exists for a time series \cite{damsleth1980interpolating,pourahmadi1989estimation,kihoro2013imputation,oh2015multiple}, it is inherently challenging to devise such a model, especially within the biomedical domain. With respect to these characteristics, {\em phenomenologically} three main types of commonly encountered ``missing values'' for univariate time series can be identified. 
The first type is {\em x-missing}, as exemplified by the photoplethysmogram (PPG) signal in Fig. \ref{fig:miss_types}(a). Mathematically, a signal $f(t)$ is x-missing if observed as $f(t)(1-\chi_I(t))$, where $I\subset \mathbb{R}$ is the {\em missing value interval(s)} and $\chi$ is the indicator function. The ``forecasting'' problem aligns with x-missing signals; that is, when $I=(T,\infty)$, with $T$ denoting the present sampling time.
The second type is {\em y-missing}, as illustrated by the airflow signal in Fig. \ref{fig:miss_types}(b). A signal $f(t)$ is y-missing when observed as, for example, $\max\{\min\{f(t),\, M\},\,-N\}$, where $M,N>0$ usually come from arithmetic underflow. This case is sometimes interchangeably called {\em oversaturation}. 
The third type is {\em corruptive}, demonstrated with the airflow signal shown in Fig. \ref{fig:miss_types}.
A signal $f(t)$ is corruptive if the observation is $f(t)(1-\chi_I(t))+(cf(t)+g(t))\chi_I(t)$, where 
$c\geq 0$ quantifies signal scaling and $g$ is a function containing information irrelevant to our interest. Depending on the scenario, $g$ could be stochastic {\em noise} or deterministic {\em artifact}. 
While the corrupted signal includes the commonly treated ``noisy signal'', we do not center on this specific case. Instead, our focus lies on extreme corruption, when $c = 0$ and $g$ can take any form.

Although in this work we focus on the aforementioned classification instead of the one in \cite{rubin1976inference}, we shall acknowledge their relevance. For example, in biomedical signals, the MCAR case can manifest when a sensor experiences random disconnections, yielding an x-missing signal (see Fig. \ref{fig:miss_types}(a)). Depending on the situation, corruptive signals caused by motion artifacts or interferences (see Fig. \ref{fig:miss_types}(c)) might be MAR or MCAR. The y-missing signal depicted in Fig. \ref{fig:miss_types}(b) can be viewed as MNAR.

\subsection{Existing approaches}

To our knowledge, there is limited research focused on handling y-missing situations. Typically, addressing y-missing data requires additional information. In cases of signals corrupted by artifacts, it is possible to convert them into x-missing configurations. This conversion involves identifying low-quality signal segments and treating them as missing values (assigned as, e.g., 0). Therefore, we will primarily discuss the x-missing scenario from this point onward.

Two common strategies are used to address x-missing signals. The first is the cut-and-stitch approach, which involves removing data at missing indexes and analyzing the signal with the available information. While simple, this method can result in the loss of important temporal details. The second method involves imputing values into the gaps. Various imputation algorithms aim to generate new data that fits the underlying dynamics, which is crucial when missing segments contain critical information. These approaches range from basic mean replacement to more complex model-based solutions. Notably, forecasting, which is similar to a specialized x-missing issue, can also be used as an imputation method.
Available techniques for non-stationary signals span genetic algorithm~\cite{lobato2015multi}, interpolation~\cite{dhevi2014imputing}, fuzzy algorithm~\cite{amiri2016missing}, manifold learning~\cite{papaioannou2022time}, Bayesian estimation~\cite{west2006bayesian}, autoregressive model based approach~\cite{sridevi2011imputation}, empirical mode decomposition (EMD)  \cite{moghtaderi2012gap,sidekerskiene2016reconstruction}, and deep learning approach like BRITS~\cite{cao2018brits}, GP-VAE~\cite{fortuin2020gp} and imputeGAN~\cite{qin2023imputegan}, among others.  
The main difference between the traditional algorithms and the deep learning methods is that the former can be applied directly to target signals without requiring lengthy training periods and large datasets.

\subsection{Our contribution}

We introduce a novel algorithm, named Harmonic Level Interpolation (\textsf{HaLI}), for handling x-missing data imputation in non-stationary signals using the adaptive non-harmonic model (ANHM). \textsf{HaLI} consists of two main parts. The first part utilizes an existing imputation technique designed for oscillatory signals to initially fill in data gaps. See Table \ref{tab:imp_methods} for a list of considered techniques and Section \ref{sec: existing methods} for a summary of these methods. In the second phase, \textsf{HaLI} performs harmonic decomposition on the imputed signal and subsequently interpolates the decomposed harmonics. Our results demonstrate improved performance, particularly when the amplitude and phases of the signal's harmonic components follow specific regularity conditions, across synthetic and real-world biomedical signals.

It is important to note the presence of analogous imputation algorithms rooted in non-stationary signal decomposition. For instance, in \cite{moghtaderi2012gap}, a gap-filling technique based on decomposing the signal into intrinsic mode functions (IMFs) by EMD is proposed. This method takes advantage of the slow oscillations and regularity of the IMFs to improve missing data imputation. In \cite{sidekerskiene2016reconstruction}, authors employ polynomial interpolation to fill in missing data on the IMF level, subsequently reconstructing the original signal through mode superposition. The advantage of EMD-based decomposition approach over other techniques was only demonstrated for short intervals of missing data. It is also worth noting that EMD-based techniques might face challenges like \emph{mode mixing} inherent to EMD. Furthermore, their theoretical underpinnings remain incomplete.

\subsection{Paper organization}
The subsequent sections of this paper are structured as follows. Sec. \ref{sec:math} describes the ANHM, a phenomenological model for non-stationary signals.  
Sec. \ref{sec:proposal} introduces our main algorithm, \textsf{HaLI}. Sec. \ref{sec: existing methods}, deferred to the Supplemental Material, summarizes existing time-series imputation methods, along with considerations for automating parameter tuning for these methods. 
In Sec. \ref{sec:exps}, the experiments on synthetic and real-world signals are described. Finally, the relevant results of our proposal and further work are discussed in \ref{sec:conclusions}. 

\section{Model of Non-Stationary Oscillatory Signals}
\label{sec:math}

Biomedical signals often exhibit dynamic traits owing to the ever-changing nature of their underlying systems. In this study, we focus on \emph{oscillatory} signals with time-varying amplitude, called amplitude modulation (AM), and time-varying frequency, called frequency modulation (FM), along with oscillatory patterns. 
Vital (patho-)physiological insights for medical diagnosis and monitoring are embedded within AM, FM, and oscillatory patterns. Previous works, such as \cite{daubechies2011synchrosqueezed,degottex2012full}, introduced the {\em adaptive harmonic model} (AHM) to capture AM and FM.
Expanding further to encompass non-sinusoidal oscillatory patterns, the AHM was broadened into the ANHM \cite{wu2013instantaneous}. Furthermore, the gradual trends often carry critical information. For instance, the mean arterial blood pressure (MAP) relies on the average arterial blood pressure (ABP) signal across a cycle. The MAP's temporal progression can be modeled through the ABP signal's trend. Consequently, signal estimation methods, including missing data imputation, must consider this gradual trend. Under this consideration, the ANHM with a {\em fixed} oscillatory pattern satisfies
\begin{equation}
	\label{eq:ANH}
	x_0(t) = \sum_{k=1}^K A_k(t)s_k(\phi_k(t)) + T(t)+\Phi(t),
\end{equation}
where $K\in \mathbb{N}$ is the number of oscillatory components in the signal $x$, $A_k\in C^1(\mathbb{R})$ is a positive function, $\phi_k\in C^2(\mathbb{R})$ is a monotonically strictly increasing function, $s_k$ is a $1$-periodic smooth function with unit $L^2$ norm, mean $0$ and $|\hat{s}_k(1)|>0$, $T(t)$ is a slow-varying trend function, and $\Phi$ is a random process with mean 0 and finite variance modeling the noise. 
We call $A_k(t)s_k(\phi_k(t))$ the $k$-th {\em intrinsic mode type} (IMT) function,  $A_k(t)$ ($\phi_k(t)$, $\phi_k'(t)>0$ and $s_k$ respectively) the AM (phase, instantaneous frequency (IF) and {\em wave-shape function} (WSF) respectively) of the $k$-th IMT function. 
To avoid distraction, we focus on the assumption that $\phi_1'(t)<\phi_2'(t)<\ldots<\phi_K'(t)$ and $\underset{t}{\inf}\ \phi_1'(t)>0$; that is, we do not consider the mode-mixing setup where the IF of different IMT functions might overlap. To guarantee the identifiability of the model \cite{chen2014non}, the following slowly time-varying conditions are needed for $A_k(t)$ and $\phi_k'(t)$:
\begin{flalign*}
	\text{(C1)} && |A_k'(t)|<\epsilon|\phi_k'(t)| &&\\
	\text{(C2)} && |\phi_k''(t)|<\epsilon|\phi_k'(t)| &&
\end{flalign*}
for all $t\in\mathbb{R}$, where $k=1,\dots,K$ and $\epsilon>0$ is a small constant. 
Moreover, $T(t)$ is assumed to vary slowly so that its spectrum is supported in $[-\delta,\delta]$, where $0\leq \delta<\underset{t}{\inf}\ \phi_1'(t)$.

In \eqref{eq:ANH}, each IMT function oscillates with a single WSF, indicating a consistent oscillatory pattern over time. Nonetheless, research highlights that in real-world signals, especially biomedical ones, the oscillatory pattern often varies rather than remaining static  \cite{lin2018wave,alian2023amplitude}. Thus, the model \eqref{eq:ANH} is extended to encompass this dynamic oscillatory pattern, also known as a {\em time-varying WSF}~\cite{lin2018wave}:
\begin{equation}
	\label{eq:ANH2}
	x_1(t) = \sum_{k=1}^K \sum_{\ell=1}^{\infty} B_{k,\ell}(t)\cos(2\pi\phi_{k,\ell}(t)) + T(t)+\Phi(t)\,,
\end{equation}
with the following assumptions satisfied for a fixed small constant $\epsilon'\geq0$:
\begin{itemize}
	\item [{(C3)}] $B_l\in C^1(\mathbb{R})\cap L^\infty(\mathbb{R})$ for $l=1,2,\ldots$ and $B_1(t)>0$ and $B_l(t)\geq 0$ for all $t$ and $l=2,3\ldots$. $B_l(t) \leq c(l) B_{1}(t)$, for all  $t\in \mathbb{R}$ and $l = 1,2,\dots$, and with $ \{c(l)\}_{l = 1}^{\infty}$ a non-negative $\ell^{1}$ sequence. Moreover, there exists $N\in \mathbb{N}$ so that $\sum_{l=N+1}^\infty B_l(t)\leq \epsilon' \sqrt{\sum_{l=1}^\infty B_l^2(t)}$ and $\sum_{l=N+1}^\infty lB_l(t)\leq D\sqrt{\sum_{l=1}^\infty B_l^2(t)}$ for some constant $D>0$.
	
	\item [{(C4)}] $\phi_l\in C^2(\mathbb{R})$ and $|\phi'_l(t) - l\phi'_1(t)| \leq \epsilon' \phi'_1(t)$, for all $t\in \mathbb{R}$ and  $l = 1,\dots,\infty$.
	
	\item [{(C5)}]  $|B'_{l}(t)| \leq \epsilon' c(l) \phi'_{1}(t)$ and $|\phi''_{l}(t)| \leq \epsilon' l \phi'_{1}(t)$ for all $t \in \mathbb{R}$, and $\sup_{l;\,B_l\neq 0}\|\phi''_l\|_\infty=M$ for some $M\geq 0$.
	
\end{itemize}
We also call \eqref{eq:ANH2} the ANHM model and $\sum_{\ell=1}^\infty B_{k,\ell}(t)\cos(2\pi\phi_{k,\ell}(t)) $ the $k$-th IMT function. This model introduces a level of regularity to amplitudes and phases (and, in turn, frequencies), allowing their estimation through smooth curve interpolation methods (see Sec. \ref{sec:step3}).
We call $B_{k,\ell}(t)\cos(2\pi\phi_{k,\ell}(t))$ the {\em $\ell$-th harmonic} of the $k$-th IMT function (with $\ell=1$ being the {\em fundamental component}). Here, $B_{k,\ell}(t)$ and $\phi_{k,\ell}(t)$ denote the associated amplitude and phase functions. An algorithm that transforms $x_1(t)$ into $\{B_{k,\ell}(t),\,\phi_{k,\ell}(t)|\, k=1,\ldots, K,\, \ell=1,\ldots,\infty\}$ alongside $T(t)$ is referred to as a {\em harmonic decomposition} algorithm. 
This model captures the time-varying WSF in the following sense. Note that \eqref{eq:ANH} can be rewritten as $x_0(t) = \sum_{k=1}^K \sum_{\ell=1}^\infty [A_k(t) a_{k,\ell}]\cos(2\pi[\ell\phi_{k}(t)+b_{k,\ell}]) + T(t) + \Phi(t)$, where $a_{k,1}>0$, $a_{k,\ell}\geq 0$ and $b_{k,\ell}\in[0,1)$ come from the Fourier series expansion of $s_k$. The time-varying WSF is captured by generalizing $a_{k,\ell}$ and $b_{k,\ell}$ to be time-varying via setting $B_{k,\ell}(t):=A_k(t) a_{k,\ell}$ and $\phi_{k,\ell}(t):=\ell\phi_{k}(t)+b_{k,\ell}$ and generalizing the associated conditions. 

In many practical applications, the WSF is not spiky and can be well modeled by finite harmonics. In this study, we focus on the following simplified model of $x_1(t)$:
\begin{equation}
	\label{eq:ANH2D}
	x(t)=\sum_{k=1}^K\sum_{\ell=1}^{D_k} B_{k,\ell}(t)\cos(2\pi\phi_{k,\ell}(t)) + T(t)+\Phi(t),
\end{equation}
where $D_k\in \mathbb{N}$ is called the {\em harmonic degree} of the $k$-th IMT function. To estimate $D_k$, we will employ trigonometric regression model selection criteria~\cite{wang1993aic,eubank1990curve}, which recent studies have shown to be effective in accurately determining the necessary number of harmonic components to capture non-sinusoidal oscillatory patterns in non-stationary signals~\cite{ruiz2022wave}. From now on, we focus on the ANHM \eqref{eq:ANH2D}.

In practice, the signal adhering to ANHM is observed over a finite interval $[0,T_s]$, where $T_s>0$, and the signal is uniformly discretized using a sampling rate $f_s = 1/\Delta t$, where $\Delta t>0$ represents the sampling period. The sampled signal is denoted as $\mathbf{x}(n) = x(n\Delta t)$ and $t_n = n\Delta t$, where $n = 1,\dots,N$. 
Our focus centers on datasets with x-missing values. The indices of signal $\mathbf{x}$ are partitioned into two subsets: the observed data subset $\mathcal{O}\subset \{1,\ldots,N\}$ and the missing data subset $\mathcal{M}\subset \{1,\ldots,N\}$. The set $\mathcal{M}$ corresponds to the indexes $n\in\{1,\ldots,N\}$ such that the signal $\mathbf{x}(n)$ is given the value $\alpha\in \mathbb{C}$, $\text{NaN}$ or any missing data symbol chosen by the user. The objective is to estimate (impute) the values $\mathbf{x}(n)$ for $n\in \mathcal{M}$.

\section{Proposed Algorithm: Harmonic Level Interpolation (\textsf{HaLI})}
\label{sec:proposal}

\begin{figure*}
	\centering
	\includegraphics[trim=0 10 0 0, width=\textwidth]{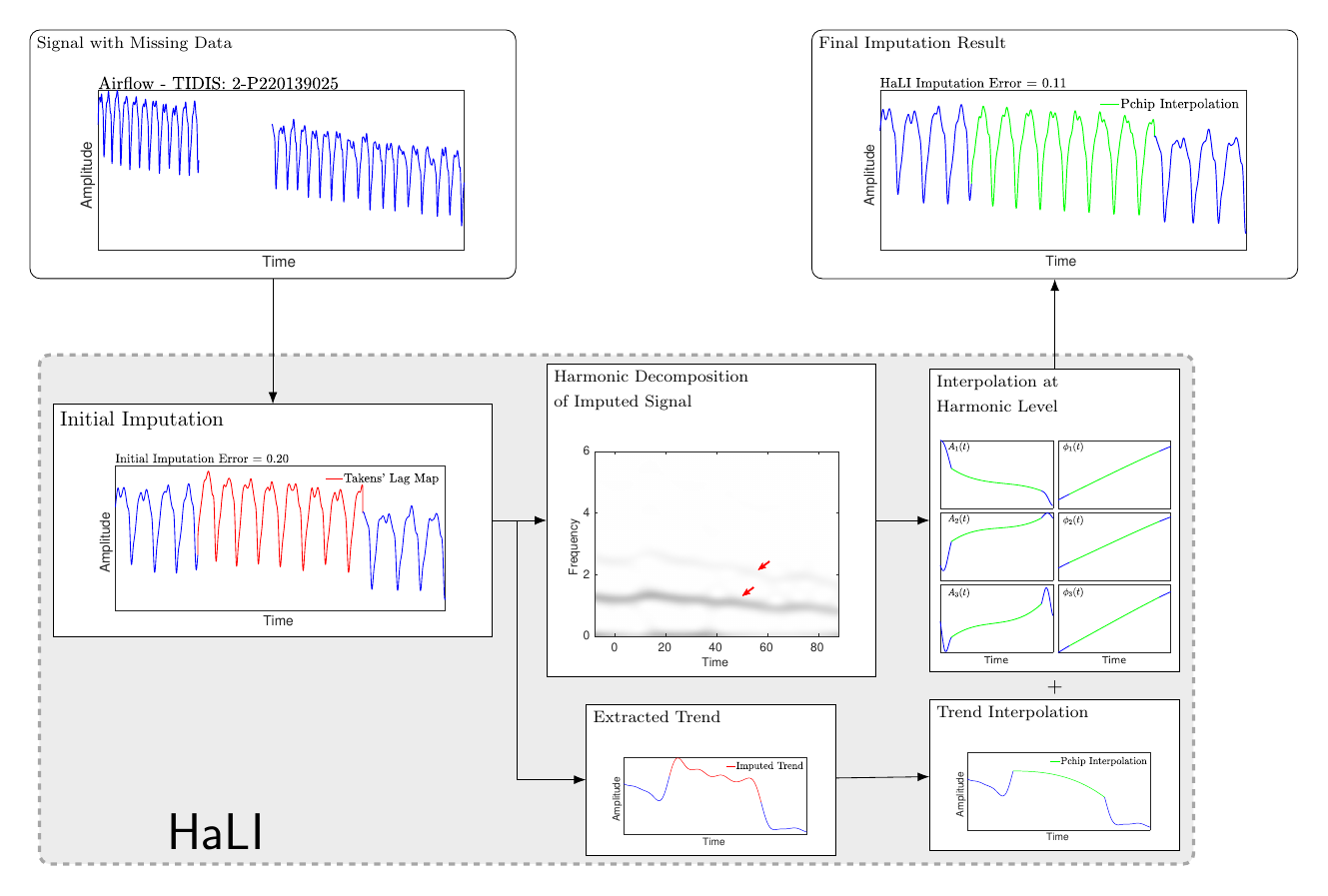}
	\caption{Proposed missing data imputation method based on interpolation of the harmonic decomposition of non-stationary signals. The three steps of the method are initial missing data imputation, harmonic decomposition and trend extraction of the imputed signal, and interpolation of the harmonic amplitude, phases, and trend to obtain the final imputation result. 
	}
	\label{fig:method}
\end{figure*}

The proposed imputation algorithm, called Harmonic Level Interpolation $\left(\textsf{HaLI}\right)$, involves three main steps. First, an initial imputation uses an existing method. Second, harmonic amplitudes and phases are acquired by decomposing the imputed signal through a time-frequency (TF) analysis or other suitable algorithm. Lastly, refined imputation occurs by interpolating the harmonic amplitudes and phases. See Fig. \ref{fig:method} for an overall flowchart of \textsf{HaLI}.  Note that \textsf{HaLI} shares a high-level idea with MissForest \cite{stekhoven2012missforest} or MISE \cite{van2011mice} for generic data, where initial imputation involves mean or median filling, followed by iterative machine learning-based imputation. \textsf{HaLI}, however, is tailored for non-stationary oscillatory time series.

Before delving into the specifics, it is essential to address some technical considerations. The initial imputation holds critical importance for \textsf{HaLI}. This step is pivotal because harmonic decomposition is sensitive to the boundary effect inherent in TF analysis tools.  
Specifically, for the x-missing signal $f(t)(1-\chi_I(t))$, due to the discontinuity and irregularity on the boundary of $I$, the TF representation is impacted near the boundary, which leads to a low-quality harmonic decomposition.
While the possibility of bypassing the initial imputation exists if a harmonic decomposition algorithm immune to this boundary effect were available, such an algorithm has not been developed to our knowledge. By employing an initial imputation, the boundary effect is alleviated and we can achieve a reliable harmonic decomposition, particularly near the boundary of $I$. Nonetheless, harmonics over $I$ hinge extensively on this initial imputation, which is usually limited. The core innovation of the proposed \textsf{HaLI} algorithm lies in enhancing imputation quality over $I$ by properly interpolating harmonics from $\mathbb{R}\backslash I$ into $I$. The subsequent sections provide a comprehensive breakdown of each step.

\subsection{Step 0: Preprocessing}
Given a x-missing signal $\mathbf{x}$, the first task is identifying the intervals containing missing values, unless they have been explicitly provided. Each detected missing value interval is described by its initial index $i_{ms}$ and the interval length $L$.
Note that in cases where the missing data symbol $\alpha$ is a finite number, we can designate a minimum interval length $I_{-}$. This allows us to consider any sequence of $\alpha$ longer than $I_{-}$ as a missing interval, encompassing even the natural $\alpha$-crossings of the signal.

\subsection{Step 1: Initial Data Imputation}
\label{sec:step1}

The first step of \textsf{HaLI} involves established techniques for handling missing data in time series. This step takes the incomplete data signal $\mathbf{x}$ and missing value intervals as input. The methods explored in this study are summarized in Table \ref{tab:imp_methods}. Each method considered may require specific parameters for model fitting, which can be predefined based on prior signal information or defaulted to preset values. Denote the resulting signal with imputed values as $\mathbf{x}_1\in \mathbb{R}^N$. Further details on these approaches are available in Section \ref{sec: existing methods}.

Among various algorithms, we emphasize Takens' embedding or Takens' Lag Map (TLM). This method consistently outperforms others, as demonstrated in Section \ref{sec:exps}. The strategy aims to locate a segment in the signal, without missing values, that closely matches the ``expected data'' in the missing data interval. The choice of a suitable similarity measure is crucial for effective results, enabling the identification of a segment with values that can be used to impute the missing data.

Before detailing the procedure, recall that Takens' embedding theorem \cite{takens1981detecting} states that under mild conditions, the properties of the underlying dynamic system generating the time series $y:\mathbb{Z}\to \mathbb{R}$ can be represented by an embedding of the time-lagged segments of the time series. Specifically, let the delay vector be $\mathbf{y}(k) = \left[y(k), y(k-\tau), \dots,y(k-(d_e - 1)\tau)\right]^\top\in \mathbb{R}^{d_e}$, where $d_e\in \mathbb{N}$ is the embedding dimension and $\tau>0$ is the lag time. With mild conditions on the underlying manifold that hosts the dynamics, $\tau$ and $d_e$, the set $\{\mathbf{y}(k)\}_k$ is diffeomorphic to the underlying manifold. In this sense, the underlying dynamic system is recovered.

Fix a missing data interval with the initial index $i_{ms}$ and the interval length $L$. Define the signal pattern to the left and the right of the missing data interval as $\mathbf{x}_{l} = [\mathbf x(i_{ms}-d),\dots,\mathbf x(i_{ms}-1)]$ and $\mathbf{x}_{r} = [\mathbf x(i_{ms}+L+1),\dots,\mathbf x(i_{ms}+L+d-1)]$, respectively. Then, we concatenate $\mathbf{x}_{l}$ and $\mathbf{x}_{r}$ to form the reference pattern $\mathbf{x}_{\texttt{tmp}} = [\mathbf{x}_{l}\ \mathbf{x}_{r}]$.  
To estimate the missing portion of the signal, a sliding window of length $2d+L$ is run over the signal outside the missing data interval, and the best imputation candidate $\mathbf{x}_p = [\mathbf x(p),\dots,\mathbf x(p+L-1)]$ is the one with the minimal Euclidean distance between $\mathbf{x}_{\texttt{tmp}}$ and the vector $[\mathbf{x}_{p,l}\ \mathbf{x}_{p,r}]$, where $\mathbf{x}_{p,l} = [\mathbf x(p-d),\dots,\mathbf x(p-1)]$ and $\mathbf{x}_{p,r} = [\mathbf x(p+L+1),\dots,\mathbf x(p+L+d-1)]$. Then, repeat the same procedure for all missing value intervals.
This way, the imputed data is chosen so that the local behavior of the signal around the candidate values is similar to the behavior around the missing data interval. From the technical perspective, we set $d_e=d$ and $\tau = 1$ for the Takens' embedding, and find ``neighbors'' over a projected subspace. We shall mention that TLM is similar to the KNNimpute proposed in \cite{troyanskaya2001missing} for DNA microarrays. 

\begin{table}[thb!]
	\small
	\centering
	\caption{Initial imputation methods considered in this study}
	\begin{tabular}{ll}
		\toprule
		Name & Reference  \\
		\midrule
		Takens' Lag Map (TLM) & \cite{takens1981detecting}  \\

		Least Square Estimation (LSE) & \cite{meynard2021efficient}  \\

		Dynamic Mode Decomposition (DMD) & \cite{schmid2010dynamic} \\

		Extended Dynamic Mode Decomposition  (EDMD) & \cite{williams2015data} \\

		Gaussian Process Regression (GPR) & \cite{rasmussen2003gaussian} \\

		ARIMA Regression with forward forecasting & \cite{box2015time}  \\

		ARIMA Regression with backward forecasting & \cite{box2015time}  \\

		Trigonometric Box-Cox, ARMA and Seasonal   & \cite{de2011forecasting} \\ 
		\hspace{10pt} Modeling (TBATS)&\\

		Sparse TF Non-linear Matching Pursuit (STF) & \cite{hou2013data}  \\

		Locally Stationary Wavelet Process (LSW) & \cite{killick2023automatic}  \\
		\bottomrule
	\end{tabular}
	\label{tab:imp_methods}
\end{table}

\subsection{Step 2: Trend Separation and Harmonic Decomposition of Imputed Signal}
\label{sec:step2}
Once the initial imputation result $\mathbf{x}_1$ is obtained, we apply a harmonic decomposition algorithm to decompose $\mathbf{x}_1$ into its trend and harmonics. There are several harmonic decomposition algorithms. However, to avoid distracting from the focus of this paper, we apply the short-time Fourier transform (STFT) based algorithm. Denote the STFT of $\mathbf{x}_1$ by $\mathbf{F}\in \mathbb{C}^{N\times N}$, using a Gaussian window $\mathbf{g}$ satisfying $\mathbf{g}(n) = e^{-\sigma n^2}$, where $\sigma>0$ is the window size. This parameter is chosen following the rule of thumb that the temporal support of $\mathbf{g}$ contains about $5\sim 8$ cycles of the oscillatory component of interest. The average cycle length is estimated locally from the average period $\tilde{T}$ of $\mathbf{x}_1$.
With $\mathbf{F}$, we find the fundamental ridge associated with each IMT function. To do this, we first compute the de-shape STFT \cite{lin2018wave}, denoted as $\mathbf{W}\in \mathbb{R}^{N\times N}$, to obtain the IF of each IMT function in $\mathbf{x}_1$. We find the ridges of $|\mathbf{W}|^2$ using a greedy ridge detection procedure~\cite{meignen2012new}. This algorithm outputs the set of fundamental ridges $\{\mathbf{c}^*_k(n)\}_{k=1}^K$, where $\mathbf{c}^*_k(n)\approx \boldsymbol{\phi}_{k,1}(n)$. Due to the continuity assumption of IFs, and hence the associated ridges, this algorithm limits the search region for each time step to a frequency band $\operatorname{FB}_{n,k} = [\mathbf{c}_k(n-1)-\operatorname{FB},\mathbf{c}_k(n-1)+\operatorname{FB}]$ around the ridge estimation at the previous time, where $\operatorname{FB}>0$ defines the maximum frequency jump for consecutive steps. The maximum frequency jump $\operatorname{FB}$ is set to $10f_s/N$. 

Based on the estimated fundamental ridges $\{\mathbf{c}_{k}^*(n)\}_{k=1}^K$, we perform the detrending of the signal by subtracting the trend estimate $\hat{\mathbf{T}}_1$ from $\mathbf{x}_1$, where
\begin{equation}
	\hat{\mathbf{T}}_1(n) = \frac{2}{g(0)}\Re{\sum_{1\leq j<\underline{\mathbf{c}}^*-\Delta} \mathbf{F}(n,j)},
\end{equation}
Here, $\underline{\mathbf{c}}^* = \underset{k}{\min}\ \mathbf{c}_k^*(n)$, $\Re{}$ means taking the real part of the complex number, and $\Delta$ is chosen close to the measure of the half-support of $\hat{\mathbf{g}}(k) = \mathcal{F}(\mathbf{g}(n))$, the Fourier transform of the analysis window.

For the harmonic decomposition, we recall that the complex fundamental component of the $k$-th IMT function of $\mathbf{x}_1$ can be recovered by integrating the STFT around the ridge $\mathbf{c}_k^*$:
\begin{equation}
	\label{eq:harm1}
	\textbf{y}_{k,1}(n) = \frac{1}{\textbf{g}(0)}\sum_{|j-\textbf{c}_{k}^*(n)|<\Delta} \textbf{F}(n,j),
\end{equation}
where $n=1,\ldots,N$. Then, the fundamental amplitude is estimated as $\tilde{\textbf{A}}_{k,1}(n) = |\textbf{y}_{k,1}(n)|$ and the associated phase, denoted as $\tilde{\boldsymbol{\phi}}_{k,1}$, is estimated by unwrapping the phase of $\textbf{y}_{k,1}$.
With the fundamental component, we extract the higher harmonics of each IMT function. To do this, for $1\leq k \leq K$, we obtain the estimate of $D_k$, denoted as $D_k^*$, using the model selection criteria based on trigonometric regression~\cite{ruiz2022wave}. After determining $D_k^*$, we find the ridge of the $\ell$-th harmonic of the $k$-th component $\mathbf{c}_{k,\ell}(n)$ for $2\leq\ell\leq D_k^*$. Given condition (C3) in Sec. \ref{sec:math}, we know that $\mathbf{c}_{k,\ell}(n) \approx \ell \mathbf{c}_k(n)$. The proper harmonic ridge $\mathbf{c}_{k,\ell}^*(n)$ is found by searching the maximum energy ridge in the vicinity of $\ell\mathbf{c}_k^*(n)$. 
Finally, the amplitude modulation of the $\ell$-th harmonic is estimated as $\tilde{\mathbf{A}}_{k,\ell}(n) = |\mathbf{y}_{k,\ell}(n)|$ for $n=1,\ldots,N$ and its phase, denoted as $\tilde{\boldsymbol{\phi}}_{k,\ell}$, is estimated by unwrapping the phase of $\textbf{y}_{k,\ell}$, where $\mathbf{y}_{k,\ell}(n)$ is the $\ell$-th complex component of $k$-th component of $\mathbf{x}_1$ obtained using \eqref{eq:harm1} by replacing $\mathbf{c}_{k}^*(n)$ for $\mathbf{c}_{k,\ell}^*(n)$. 

Note that we could employ advanced TF analysis algorithms like the synchrosqueezing transform \cite{daubechies2011synchrosqueezed}, newer ridge detection algorithms \cite{Colominas_Meignen_Pham_2020,Laurent_Meignen_2021}, or sophisticated reconstruction formulas at this stage. However, for the sake of algorithm simplicity, computational efficiency, and accuracy, we choose to follow this pragmatic approach.

\subsection{Step 3: Interpolation at the Harmonic Level}
\label{sec:step3}

The harmonic decomposition of $\mathbf{x}_1$ can enhance imputation by employing amplitude and phase clipping, followed by interpolation within the missing data interval(s). Various curve interpolation schemes are applicable in this stage. However, due to the regularity of amplitude and phase functions, we anticipate smoother curves with minimal oscillations. We consider two effective yet simple interpolation techniques: cubic spline interpolation and shape-preserving cubic Hermite interpolation (pchip) \cite{fritsch1980monotone}. It is worth noting that splines tend to overshoot within the data gap, while pchip produces smoother curves with fewer oscillations. The specific procedure for this step is detailed below.

\begin{enumerate}
	\item The harmonic amplitudes $\tilde{\mathbf{A}}_{k,\ell}$ and phases $\tilde{\boldsymbol{\phi}}_{k,\ell}$ are clipped in the missing data interval $I_{\texttt{ms}}$.
	\item The amplitude and phases are interpolated on $I_{\texttt{ms}}$, resulting in interpolated amplitude estimate $\overline{\mathbf{A}}_{k,\ell}$ and interpolated phase estimate $\overline{\boldsymbol{\phi}}_{k,\ell}$.
	\item The final imputation result $\mathbf{x}_2$ is constructed by harmonic superposition in the missing data interval:
	\begin{equation*}
		\mathbf{x}_2(n) := \sum_{k=1}^K \sum_{\ell=1}^{D_k^*} \overline{\mathbf{A}}_{k,\ell}(n)\cos(2\pi\overline{\boldsymbol{\phi}}_{k,\ell}(n))
	\end{equation*}
	for $n\in I_{\texttt{ms}}$.
	The remaining part of $\mathbf{x}_2$ satisfies 
	$\mathbf{x}_2(n) := \mathbf{x}_1(n)$
	for $n\in \{1,\ldots,N\}\backslash I_{\texttt{ms}}$.
\end{enumerate}
Missing values within the trend $\hat{\mathbf{T}}_1$ are directly interpolated to produce the final estimate $\hat{\mathbf{T}}_2$. This interpolation method is chosen because time-varying trends typically exhibit smooth variations and lack local oscillations at the scale of data gaps.

The resulting output of the \textsf{HaLI} algorithm is the imputed signal $\mathbf{x}_2 + \mathbf{\hat{T}}_2$. Importantly, this final imputation result represents a denoised estimate of $\mathbf{x}$. Signal denoising is a common task in various signal-processing applications. In essence, our proposed method not only imputes missing data but also simultaneously denoise signal, a topic we will explore further in the numerical experiments section.

\section{Summary of existing imputation algorithms for time series}\label{sec: existing methods}

In this section, we review the existing imputation methods from Table \ref{tab:imp_methods}, which we use as initial imputation. We also give some guidelines regarding the tuning of their parameters.

\subsection{Algorithms}

\subsubsection{Least Square Estimation and Dynamic Mode Decomposition}

Nonlinear dynamic system identification is a common approach for time series missing data imputation. Generally, a parametric model is fit to the known data and then a forecasting procedure is used to estimate the future values of the time series. For data imputation, we consider the data points left to the missing interval as a time series to be forecasted into the missing data interval. Matrices $\mathbf{X}\in\mathbb{R}^{K\times M}$ and $\mathbf{Y}\in\mathbb{R}^{K\times M}$, where $\mathbf{X}$ contains the lagged values of the time series and $\mathbf{Y}$ contains the corresponding values to be predicted, are defined as
\begin{align}
	\label{eq:LSEX}
	\mathbf{X} =& \left[\mathbf{x}_0,\mathbf{x}_1,\dots,\mathbf{x}_{M-1}\right],\\
	\label{eq:LSEY}
	\mathbf{Y} =&  \left[\mathbf{x}_1,\mathbf{x}_2,\dots,\mathbf{x}_{M}\right]\,,
\end{align}
where $\mathbf{x}_l\in\mathbb{R}^K$ is given by $[x(i_{ms}-M-K+l),\dots,x(i_{ms}-M+l-1)]$, for $1\leq l \leq M$; that is, each column of $\mathbf{Y}$ is obtained by shifting the element of the $l$-th column of $\mathbf{X}$ by one sample. Then, a forecasting model is fit to $\mathbf{X}$ and $\mathbf{Y}$ using a least squares estimation (LSE) algorithm
\begin{equation}
	\underset{\mathbf{A}}{\min} \|\mathbf{Y}-\mathbf{AX}\| ^2,
\end{equation}
where the solution is $\tilde{\mathbf{A}} = \mathbf{Y}\mathbf{X}^T\left(\mathbf{X}\mathbf{X}^T\right)^{-1}$ if its exists.
Once the forecasting model is fit, the missing values are imputed by setting $\mathbf{x}_1(i_{ms}+m) = \tilde{\mathbf{A}}^m \mathbf{x}_{K}$, where $m$ is the imputed data index.

The dynamic mode decomposition (DMD) for system identification follows the same formulation as LSE but the matrix $\mathbf{A}$ is estimated from the singular value decomposition of the data matrix $\mathbf{X}$.

\subsubsection{Extended Dynamic Mode Decomposition}

Formally, this method seeks to estimate the non-linear evolution operator, denoted as $\mathbf{F}$, of a discrete-time dynamical system: \text{$\mathbf{x}(n+1) = \mathbf{F}(\mathbf{x}(n))$}. To achieve this, the extended dynamic mode decomposition (EDMD) approach approximates the infinite-dimensional \emph{linear} Koopman operator, denoted as $\mathcal{K}$, which governs the evolution of the system \emph{observables}. These observables are functions defined in the state space that define the time evolution of the system. While the Koopman operator provides a \emph{global} linearization of the dynamical system, LSE and DMD perform a \emph{local} linearization.

In~\cite{williams2015data}, the authors propose a data-driven algorithm that approximates the eigenvalues $\{\mu_j\}$ and modes $\{\mathbf{v}_j\}$ of the finite-dimensional Koopman operator estimate $\hat{\mathbf{K}}$. They demonstrate that this method extends the original DMD proposal, improving mode estimates of the Koopman operator.

Furthermore, in the study conducted by Hua et al.\cite{hua2017high}, a novel algorithm is introduced to efficiently estimate the modes. This algorithm uses a kernel function to project the data into a high-dimensional feature space, which improves the approximation of the Koopman operator. Based on this procedure, the missing data is imputed by forecasting into the missing data interval by 
\begin{equation}
	\label{eq:forEDMD}
	\mathbf{x}(n+1) = \sum_{j} \mu_k \mathbf{v}_j \phi_j(\mathbf{x}(n)),
\end{equation}
\noindent where $\{\phi_j(\mathbf{x})\}$ is a dictionary of eigenfunctions that map the observables $\mathbf{x}$ into the state space. Like with $\mathcal{K}$, the infinite-dimensional eigenfunctions are approximated by the linear combination of a set of finite-dimensional functions $\boldsymbol{\psi}_j(\mathbf{x})$ defined in the feature space of dimension $J$ which is in general much greater than both $K$ and $M$. In order to avoid a high-dimensional problem, the functions $\boldsymbol{\psi}(\mathbf{x})$ are defined implicitly by using a kernel $\mathbf{k}(\mathbf{x}_i,\mathbf{x}_j) = \langle \boldsymbol{\psi}(\mathbf{x}_i),\boldsymbol{\psi}(\mathbf{x}_j) \rangle$, which is the inner-product between two $J$-dimensional eigenfunctions.  In this work, a Gaussian kernel $\mathbf{k}(\mathbf{x}_i,\mathbf{x}_j) = \exp(-\|\mathbf{x}_i-\mathbf{x}_j\|_2/\gamma)$ is used, where the kernel size $\gamma$ is chosen by the user.

\subsubsection{Gaussian Process Regression}

Gaussian Process Regression (GPR) is a machine learning technique used for regression tasks. It models the relationship between input variables (features $\mathbf{x}$) and output variables (responses $\mathbf{y}$) in a probabilistic manner via  $\mathbf{y} = f(\mathbf{x})$. Instead of predicting a single value as in traditional linear regression, GPR provides a probability distribution over possible target values for a given input. In the imputation application, a Gaussian process (GP) is trained using the available data, which allows the trained GP to estimate missing values of a time series based on the posterior probability distributions.
To implement the GPR imputation method, we used the $\mathrm{fitrgp}$ method from the Statistical and Machine Learning Toolbox of MATLAB$^\copyright$. To train the model, we used the matrix $\mathbf{X}$ as defined in \eqref{eq:LSEX} and use the matrix $\mathbf{Y}$ as the target. This approach performs dynamic model estimation using a non-parametric method, which can be beneficial in cases where the underlying system is complex and hard to model parametrically.

\subsubsection{ARIMA Regression with forward forecasting}

ARIMA (Autoregressive Integrated Moving Average) is a widely adopted time-series modeling technique employed for forecasting future values in non-stationary time series data \cite{brockwell2009time}. Moreover, to capture periodic components within the signal, one can resort to the seasonal ARIMA (SARIMA) models \cite{brockwell2009time}. The modeling process entails the identification of underlying data patterns by examining autocorrelations and partial autocorrelations. Once these patterns are discerned, the model becomes a valuable tool for generating future values. In the context of missing data imputation, the approach involves fitting the SARIMA model to the data preceding the missing interval, followed by imputing the missing data  utilizing the fitted model. Notably, one of the pivotal parameters in this approach is the seasonability period, denoted as $\lambda$, among other parameters. 

\subsubsection{ARIMA Regression with backward forecasting}

This method is analogous to the previous method but performs a backward prediction step. First, we fit the ARIMA model to the flipped version of the data to the right of the missing interval. $\mathbf{x}_{pos} =[x(N),x(N-1),\dots,x(i_{ms}+L+1)]$. Then, the missing data interval is predicted from the fit ARIMA model and flipped into the forward direction. As in the previous method, the periodicity $\lambda$ needs to be set beforehand.

\subsubsection{TBATS Modeling}
The TBATS (Trigonometric seasonality, Box-Cox transformation, ARMA errors, Trend, and Seasonality) model is a state-space model designed for time series analysis \cite{de2011forecasting}. It incorporates exponential smoothing to give more weight to recent data during estimation. The trend and seasonal components are also smoothed using exponential methods to enhance local data estimation. Additionally, the Box-Cox transformation is applied to normalize the data, improving its statistical characteristics. ARMA errors are included to capture complex data dynamics. Similar to ARIMA models, TBATS can be employed for missing data imputation by fitting the model to the data before the missing interval and forecasting the missing values.  In \cite{de2011forecasting}, an efficient algorithm is proposed for estimating model parameters using maximum likelihood optimization on the state-space representation of the model.

\subsubsection{Data-Driven Time-Frequency Analysis}

The sparse time-frequency (TF) decomposition method \cite{hou2013data} was used for imputing signals with missing data. Their nonlinear matching pursuit algorithm decomposes non-stationary signals into phase- and amplitude-modulated harmonic components, represented as $A_k(t)\cos(\theta_k(t))$. This method actively seeks optimal amplitude and phase functions using an overcomplete Fourier dictionary while promoting sparsity through $\ell$-1 norm regularization of the amplitude coefficients. To address missing data, they employ an interior-point method based on preconditioned conjugate gradient~\cite{kim2007interior}. This algorithm requires setting specific parameters in advance, including an initial phase estimator and the number of harmonic components denoted as $K$.

\subsubsection{Locally Stationary Wavelet Process Forecasting}
Fix a mother wavelet $\psi(t)$ with a compact support. In the locally stationary wavelet (LSW) model \cite{nason2000wavelet}, a time series $\{X_{t,T}\}_{t=0,1,\dots,T-1}$, where $T \in \mathbb{N}$, is a realization of a LSW process if it satisfies 
\begin{equation}
	\label{eq:LSW}
	X_{t,T} = \sum_{j=1}^\infty\sum_{k=-\infty}^\infty w_{j,k;T}\psi_{j,k}(t)\zeta_{j,k}\,,
\end{equation}
where, $\psi_{j,k}(t)$ represents a family of discrete non-decimated wavelets translated and dilated from $\psi$, $\zeta_{j,k}$ is a set of uncorrelated random variables with mean $0$ and variance $1$, and $w_{j,k;T}$ are the amplitudes of the process satisfying some mild conditions so that $X_{t,T}$ is a locally stationary random process. 
Within the LSW framework, a set of observations $X_{0,T},\dots,X_{t-1,T}$ can be used to predict the next observation $X_{t,T}$ by a linear combination of the most recent $p$ observations, as proposed in~\cite{fryzlewicz2003forecasting}, where the weights of linear combination are chosen by minimizing the mean square prediction error (MSPE).

\subsection{Automatic imputation parameters tuning}
\label{sec:parameters}

Parameter tuning is essential and can greatly affect algorithm performance. Typically, unless an automatic algorithm is available, parameters are manually adjusted to achieve satisfactory results. For those algorithms without standard parameter selection procedure, we provide empirical data-driven criteria that can be applied to a wide range of oscillatory time series, and leave their theoretical justification to future work.

\subsubsection{TLM algorithm}
The TLM algorithm requires the template length $d$ to be set beforehand. This length defines the pattern to be matched to the rest of the signal and can significantly impact the final result. Given that we focus on the study of oscillatory signals, we propose an intuitive way to define $d$ based on the quasi-periodicity of the signal. Given the average period $\Tilde{T}$ as $\Tilde{T} = 2\pi/f_p$, where $f_p$ is the most energetic frequency of the signal that can be estimated by SST, we then set $d = k\Tilde{T}$ with $k\in\mathbb{N}$. We empirically find that $k=3$ leads to satisfactory results in most signals. Note that the slow-varying nature of the instantaneous frequency means that the signal can be locally approximated by a harmonic function with minimal loss in accuracy.

\subsubsection{Dynamic signal forecasting}
Algorithms like LSE, DMD and EDMD involve two adjustable parameters: the dimension of the embedding space denoted as $K$, which is related to the maximum shift between subsignals, and the length of the subsignals represented by $M$, used in constructing the system matrix $\mathbf{A}$. When dealing with nonstationary signals, it's essential to set $M$ to be greater than the signal's average period to capture intercycle variability effectively. It is recommended to select $M$ so that it encompasses at least three complete signal cycles within each subsignal.
The embedding dimension $K$ needs to be chosen such that it is greater than $M$. Theoretical findings in~\cite{meynard2021efficient} illustrate that as $K$ approaches infinity, the variance of the forecast result asymptotically decreases, but the bias increases. As a practical compromise, we empirically opt for $K=2.5M$ in all experiments within this study.

\subsubsection{ARIMA regression}
To fit a SARIMA model to the data, either before or after the missing interval for a good imputation, estimating the periodicity (or seasonality) $\lambda$ of the signal is crucial due to the time-varying nature of the signals of interest. However, to our knowledge, there does not exist a proper approach to estimate the seasonality of signals with time-varying frequency for SARIMA. We thus consider the following empirical approach. We first estimate the length of each cycle of the signal as $T_k = \#\{\hat\phi_1(n):\ k-1<\hat\phi_1(n)\leq k\}$, where $\hat\phi_1$ is the estimated phase and $k\in\mathbb{N}$. Since the seasonality is assumed fixed in SARIMA, we estimate the seasonality for the ARIMA forward regression method by $\lambda_f = \frac{1}{N_c}\sum_{i=a-N_c}^{i=a}T_i$, where $a$ denotes cycle immediately preceding the missing data interval. Similarly, for the ARIMA backward regression method, we estimate the seasonality by $\lambda_b = \frac{1}{N_c}\sum_{i=b}^{i=b+N_c}T_i$, where $b$ denotes the first cycle immediately after the missing data interval. In either case, $N_c$ needs to be set beforehand. 
Due to the time-varying nature of the considered model, local estimates are preferable as they more accurately capture the dynamics of the data in the vicinity of the missing interval, but $N_c$ should not be too small to avoid large estimate variation. As a general guideline, an empirical choice $N_c = 3$ typically leads to a good estimate for the missing interval. Other parameters, including the orders of the model and parameters, could be estimated following the standard procedure.  

\subsubsection{TBATS}
The general procedure for estimating the state-space model involves two main steps: parameter estimation and model selection. In the first step, optimal parameters are determined by maximizing the likelihood of the data given a specific model. In the second step, the best model is selected using an information criterion.
For the TBATS model, various configurations are tested, such as whether to apply Box-Cox transformation or include ARMA errors in the model. In the context of the signals studied in this work, special attention is given to the periodicity of the seasonal components and the smoothing parameter for the trend due to the nature of time-varying frequency. Estimating the periodicity of the fundamental component, linked to the longest seasonal component of the model, is crucial to ensure accurate forecasting of missing data. We follow the empirical estimation approach used for SARIMA models discussed in the previous section to handle the time-varying frequency issue.

\subsubsection{Data-Driven Time-Frequency Analysis}
For the initial phase estimate $\theta_0(t)$ we use the linear function $\theta_0(t) = 2\pi f_0 t$, where $f_0$ is chosen as the maximum energy frequency of the spectrum of $x(t)$. The number of harmonic components $K$ is estimated by finding the optimal order of the adaptive non-harmonic model fit to the data, as described in~\cite{ruiz2022wave}.

\subsubsection{Locally Stationary Wavelet Processes}: The number $p$ of past indexes used to estimate the forecasting coefficients is automatically estimated using the local partial autocorrelation function (LPACF) as proposed in \cite{killick2020local}. Authors in \cite{xie2009forecasting} studied the choice of mother wavelet and found that the $8$th order extremal phase Daubechies wavelet performs adequately for the majority of cases. Additionally, in order to compute $\mathbf{B}_{t,T}$, a $(t+1)\times(t+1)$ that approximates the covariance matrix of $X_{0,T},\dots,X_{t-1,T}$ used to find the minimizer of the MSPE, a smoothing of the wavelet periodogram needs to be performed. The smoothed periodogram bandwidth is estimated using the automatic procedure proposed by Nason~\cite{nason2013test}.

\section{Numerical results}\label{sec:exps}

Within this section, we present two sets of analyses to showcase the efficacy of our proposed \textsf{HaLI} algorithm. The initial experiment involves synthetic signals adhering to the ANHM, exhibiting time-varying WSF. This model is utilized to showcase our method's proficiency in restoring missing data when oscillation patterns shift within the data gap. The subsequent experiment applies our approach to real-world signals, further highlighting its enhancements over the initial imputation. Considering the multiple options for Step 1, we conduct a comparative study encompassing diverse algorithms listed in Table \ref{tab:imp_methods}.
The Matlab code of \textsf{HaLI} is available at \url{https://github.com/joaquinr-uner/MSI_HarmDecomp}, where the readers can also find codes that reproduce the reported results. Likewise, the ready-to-use package is available at  \url{https://github.com/joaquinr-uner/harmonic_imputation}.

In this section, to specify which algorithm is used in the initial and third steps of the \textsf{HaLI} algorithm, we adopt the notation $\mathsf{HaLI[IMP](int)}$, wherein $[\mathsf{IMP}]$ indicates the imputation method employed in the initial step and $\mathsf{(int)}$ represents the interpolation scheme utilized in Step 3. The chosen interpolation method is indicated by $\mathsf{int}= \mathsf{s}$ for splines or $\mathsf{int}= \mathsf{p}$ for pchip interpolation.

\subsection{Synthetic signals}
\label{sec:synth}
\begin{figure}
	\centering
	\includegraphics[trim=0 20 0 0, width=\columnwidth]{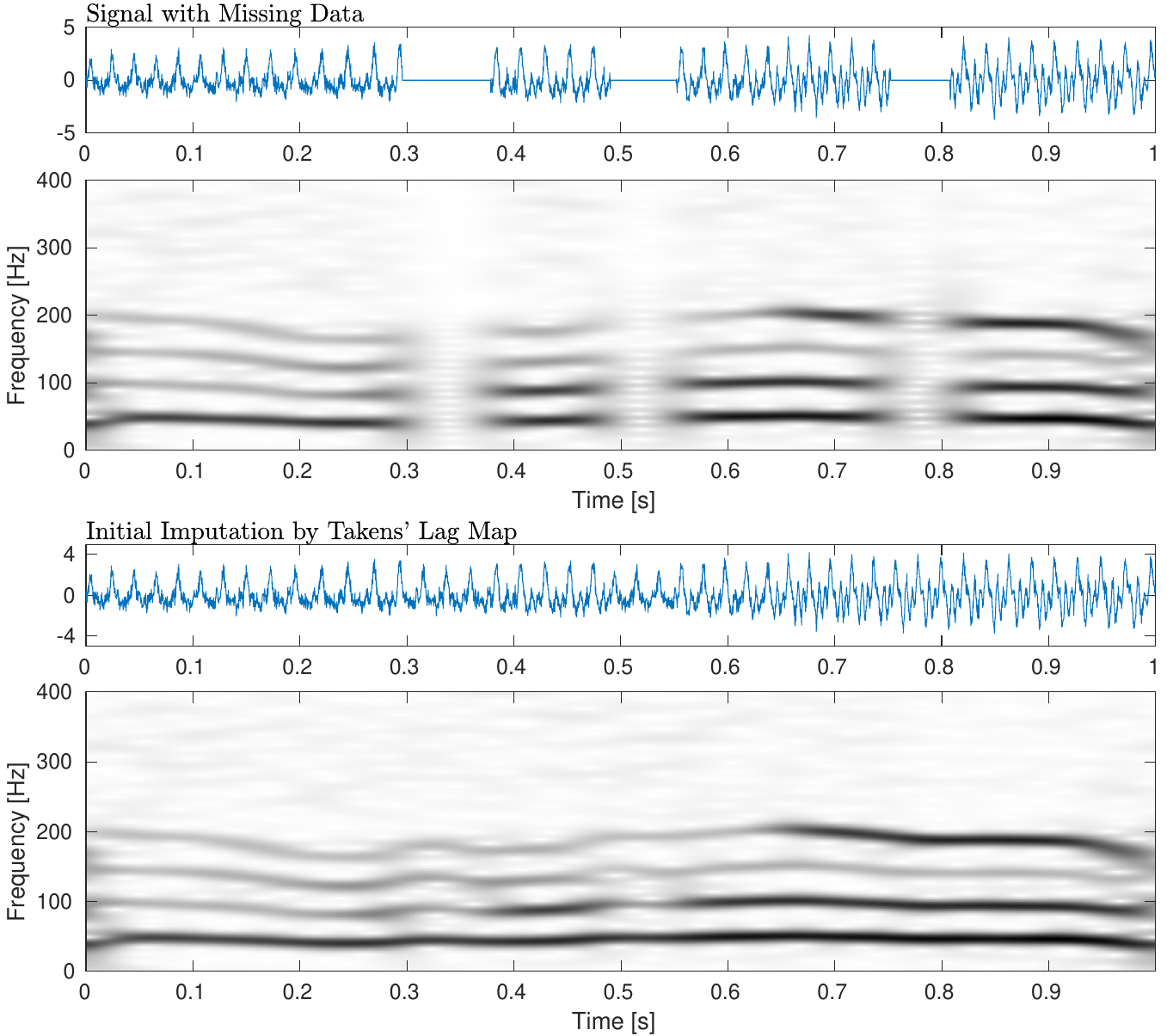}
	\caption{Top: Example noisy synthetic signal with time-varying wave-shape and $20\%$ of missing data alongside the modulus of its STFT. Bottom: Imputed version of the signal using Takens' Lag Map in the initial imputation step alongside the modulus of its STFT.}
\label{fig:missing}
\end{figure}

We consider simulated signals satisfying the ANHM \eqref{eq:ANH2D}. We generate these signals by defining the instantaneous amplitude and phase as $B_1(t) = \sqrt{t+1}$ and $\phi_1(t) = 50t + 5/(2\pi)\cos(2\pi t) + Y(t)$, respectively, where $Y(t)$ is a random process defined as $Y(t) = \int_{0}^t \frac{R_B(u)}{\|R_B(u)\|_\infty} du$, where $R_B(t)$ is a moving averaged standard Gaussian white noise. The harmonic phases are set to $\phi_\ell(t) = e_{\ell}\phi_1(t)$ for $\ell\geq 2$, where $e_{\ell}$ is a uniform random variable supported on $[0.95\ell,1.05\ell]$. The harmonic amplitudes $B_\ell(t)$ are related to $B_1(t)$ through the condition (C4) so that, for each harmonic $\ell\geq 2$, $\alpha_\ell(t) = B_\ell(t)/B_1(t)$ is slowly time-varying that is independent from the other harmonics. The signals are synthesized using the sampling rate $4000$ Hz with the duration $T_s= 1$ s. For each signal, we simulate missing values with various percentages of missingness $P_{ms} \in [0.05,0.2]$ and generate three missing data intervals with different lengths such that $L_1 + L_2 + L_3 = NP_{ms}$, where $L_i$, $i = 1, 2 ,3$, is the length of the $i$-th missing data interval. Regarding the noise component $\Phi(t)$, we considered a zero-mean Gaussian distribution with two levels of variance corresponding to $\operatorname{SNR}$ values of $10$ and $20$ dB. We also conducted runs with no added noise to establish a reference. In total, we generated $100$ signals, each with randomly placed missing values at varying levels of missing data and noise. Therefore, we processed a total of $300$ random signals for each level of missingness.
Figure \ref{fig:missing} illustrates the effect of the initial data imputation. The top two rows display the original noisy signal ($\operatorname{SNR} = 10$ dB) with $20\%$ missing data across three intervals and the modulus of its STFT ($|\mathbf{F}|$), revealing TF plane discontinuities and blurring of harmonic ridges. The third and fourth rows show the signal after initial imputation and the modulus of its STFT, where artifacts due to missing data are reduced. The \textsf{HaLI} algorithm further enhances this TFR representation through harmonic decomposition and interpolation.

\begin{figure}[t!]
	\centering
	{\footnotesize Noiseless}
	
	\includegraphics[trim=0 0 0 0, width=0.55\columnwidth]{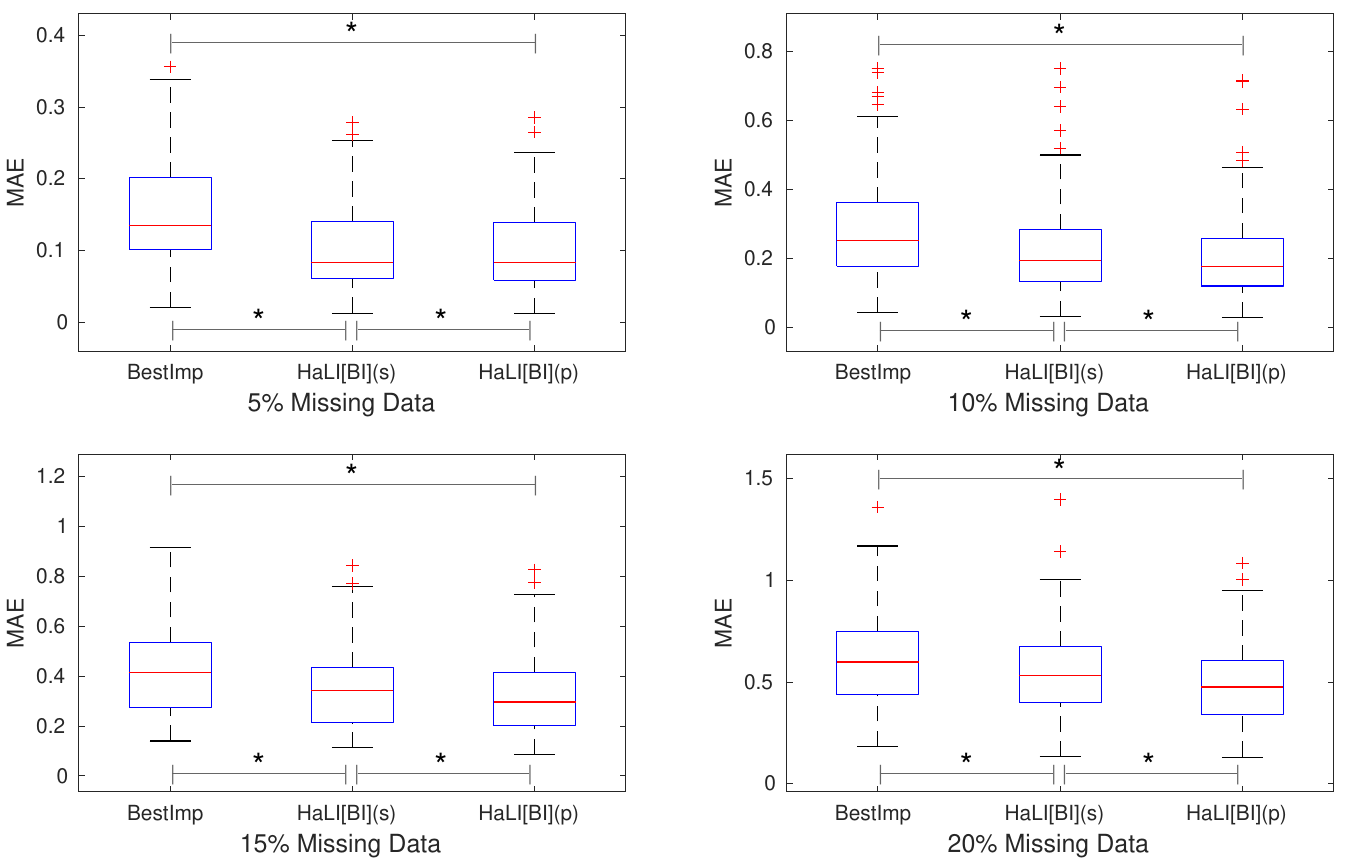}
	\vspace{.02in}
	
	{\footnotesize$\operatorname{SNR} = 20$ dB}
	
	\includegraphics[trim=0 0 0 0, width=0.55\columnwidth]{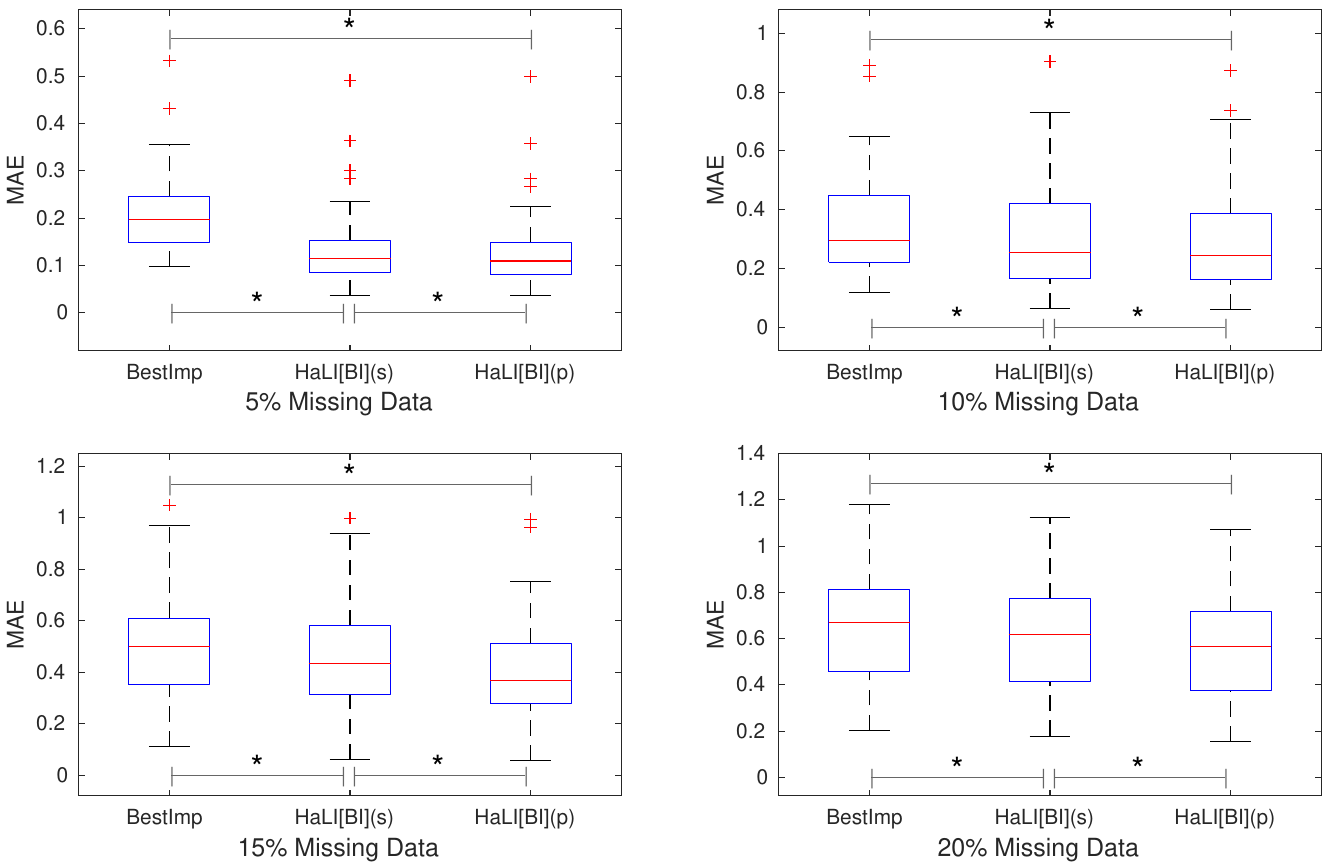}
	\vspace{.02in}
	
	{\footnotesize$\operatorname{SNR} = 10$ dB}
	
	\includegraphics[trim=0 0 0 0, width=0.55\columnwidth]{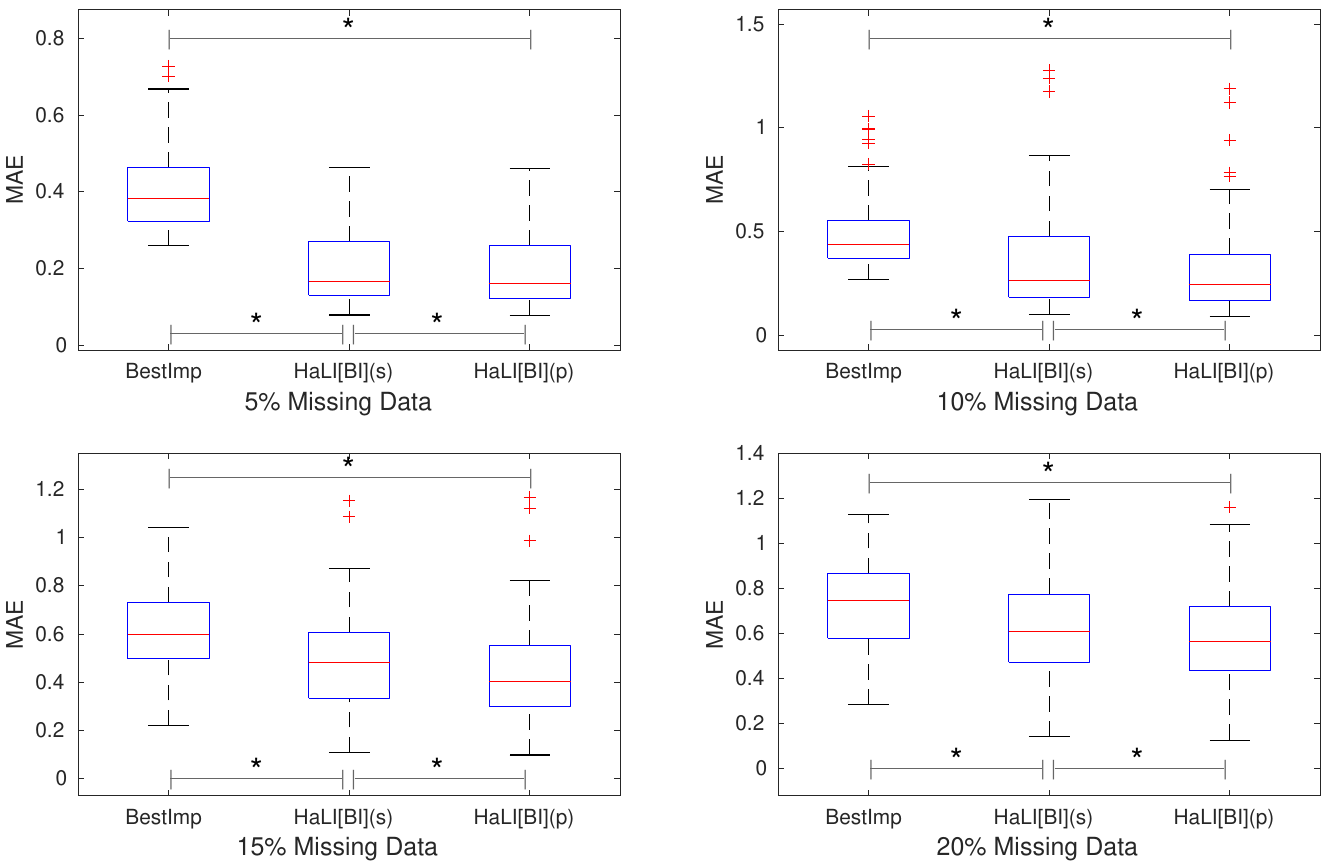}
	\caption{Boxplot of the MAE for synthetic signals at different noise levels. For each missing data rate, boxplots for the best imputation method $\left(\mathsf{BI}\right)$ and the complete $\mathsf{HaLI}$ algorithm with spline $\left(\mathsf{HaLI[BI](s)}\right)$ and pchip $\left(\mathsf{HaLI[BI](p)}\right)$ interpolation schemes. In both cases, the harmonic decomposition and interpolation steps are performed on the result of the best-performing initial imputation method based on the $\operatorname{MAE}$. Statistically significant differences ($p<0.0167$) between the median values of the three methods according to the signed-rank Wilcoxon test are indicated with `*'. }
	\label{fig:results_synth}
\end{figure}

To evaluate each imputation method, we computed the mean absolute error ($\mae$) across simulated signals. 
First, we assess various initial imputation methods detailed in Table \ref{tab:imp_methods}. 
The optimal initial imputation method is denoted as $\mathsf{BI}$ (Best Imputation). 
The $\mae$ of the final results by \textsf{HaLI} are computed and compared to those of the initial imputation. The Wilcoxon signed-rank hypothesis test with Bonferroni correction was used to determine if the differences in the median values of the $\mae$ before and after interpolation were statistically significant.  

\begin{figure}
\centering
\includegraphics[trim=0 10 0 0, width=1\columnwidth]{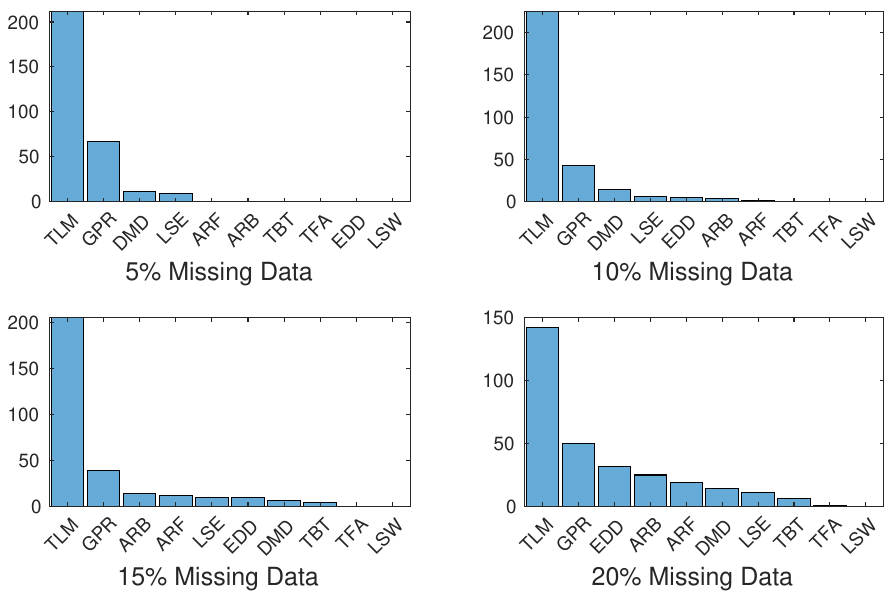}
\caption{Imputation performance on synthetic signals. Frequency histograms of the best performing initial imputation method on synthetic signals for each missingness rate. References: TLM: Takens' Lag Map. LSE: Least Square Estimation. DMD: Dynamic Mode Decomposition. EDD: Extended Dynamic Mode Decomposition. GPR: Gaussian Process Regression. ARF: ARIMA Regression with Forward Forecasting. ARB: ARIMA Regression with Backward Forecasting. TBT: Trigonometric Box-Cox, ARMA, and Seasonal Forecasting. TFA: Data-Driven Time-Frequency Analysis. LSW: Locally Stationary Wavelet Process.}
\label{fig:hist_synth}
\end{figure}

\begin{table}[h]
\centering
\caption{Median MAE values across each initial method and the \textsf{HaLI} procedure for synthetic signals at each noise level. I: Best initial imputation. S: \textsf{HaLI[BI](s)}. P:\textsf{HaLI[BI](p)}, where $\textsf{BI}$ means Best Imputation.}
\small
\begin{tabular}{|c|l|c|c|c|c|c|}
	\hline
	& \multicolumn{2}{|c|}{} & \multicolumn{4}{|c|}{$\mathbf{P_{ms} [\%]}$} \\
	\hline
	& \multicolumn{2}{|c|}{} & \textbf{5} &\textbf{10} & \textbf{15} & \textbf{20}  \\
	\hline
	\multirow{6}{*}{\rotatebox[origin=c]{90}{Noiseless}}
	& \multicolumn{2}{|c|}{$\mae_I$} & 0.1352 & 0.2505 & 0.4167 & 0.5990 \\
	
	& \multicolumn{2}{|c|}{$\mae_S$} & 0.0842 & 0.1948 & 0.3436 & 0.5304 \\
	
	& \multicolumn{2}{|c|}{$\mae_P$} & 0.0832 & 0.1768 & 0.2967 & 0.4762  \\
	
	& \multicolumn{2}{|c|}{$p_{I-S}$} & $<0.0001$  & $<0.0001$ & $<0.0001$ & $<0.0001$  \\
	
	& \multicolumn{2}{|c|}{$p_{I-P}$} & $<0.0001$ & $<0.0001$ & $<0.0001$ & $<0.0001$ \\
	
	& \multicolumn{2}{|c|}{$p_{S-P}$} & $<0.0001$ & $<0.0001$ & $<0.0001$ & $<0.0001$ \\

	\hline
	\multirow{6}{*}{\rotatebox[origin=c]{90}{$20$ dB}} &      
	\multicolumn{2}{|c|}{$\mae_I$} & 0.1971 & 0.2943 & 0.4994 & 0.6685 \\
	
	&    \multicolumn{2}{|c|}{$\mae_S$} & 0.1135 & 0.2554 & 0.4349 & 0.6172 \\
	
	&    \multicolumn{2}{|c|}{$\mae_P$} & 0.1093 & 0.2437 & 0.3713 & 0.5640 \\
	
	&    \multicolumn{2}{|c|}{$p_{I-S}$} & $<0.0001$  & $<0.0001$ & $0.0001$ & $<0.0001$ \\
	
	&    \multicolumn{2}{|c|}{$p_{I-P}$} & $<0.0001$  & $<0.0001$ & $<0.0001$ & $<0.0001$ \\
	
	&    \multicolumn{2}{|c|}{$p_{S-P}$} & $<0.0001$  & $<0.0001$ & $<0.0001$ & $<0.0001$ \\

	\hline
	\multirow{6}{*}{\rotatebox[origin=c]{90}{$10$ dB}} &
	
	\multicolumn{2}{|c|}{$\mae_I$} & 0.3818 & 0.4396 & 0.5963 & 0.7445 \\
	
	&    \multicolumn{2}{|c|}{$\mae_S$} & 0.1663 & 0.2664 & 0.4805 & 0.6094 \\
	
	&    \multicolumn{2}{|c|}{$\mae_P$} & 0.1620 & 0.2444 & 0.4052 & 0.5652 \\
	
	&    \multicolumn{2}{|c|}{$p_{I-S}$} & $<0.0001$  & $<0.0001$ & $<0.0001$ & $<0.0001$ \\
	
	&    \multicolumn{2}{|c|}{$p_{I-P}$} & $<0.0001$  & $<0.0001$ & $<0.0001$ & $<0.0001$ \\
	
	&    \multicolumn{2}{|c|}{$p_{S-P}$} & $<0.0001$  & $<0.0001$ & $<0.0001$ & $<0.0001$ \\
	\hline
\end{tabular}
\label{tab:results_synth}
\end{table}

Figure \ref{fig:results_synth} displays the experiment's outcomes through boxplots, and Table \ref{tab:results_synth} presents median $\mae$ values and corresponding $p$-values for the comparison between the best initial imputation and the complete \textsf{HaLI} algorithm using both interpolation methods. The results substantiate that the proposed $\mathsf{HaLI}$ approach consistently surpasses initial imputation across various missingness rates, interpolation tactics, and noise levels. Notably, shape-preserving cubic interpolation consistently outperforms cubic spline interpolation in all scenarios, attributed to pchip's reduced oscillations and overshooting compared to splines. Furthermore, the Wilcoxon signed-rank test validates the statistically significant enhancement in all instances when employing either spline or pchip interpolation ($p$-value $<0.0001$).

To compare the imputation strategies in the first step, we present frequency histograms for the best-performing initial imputation approach in Fig. \ref{fig:hist_synth}. We observe that the TLM algorithm consistently outperforms other methods across varying missing data rates. Among these, the dynamic signal forecasting methods (GPR, DMD, LSE, and EDMD) follow TLM for $P_{ms} = 5\%$ and $10\%$. From $P_{ms} = 15\%$ onwards, both ARIMA methods demonstrate similar performance, consistently outperforming other methods except TLM and GPR. At $P_{ms} = 20\%$, TLM is still the most frequent best-performing method, followed by GPR and EDMD. Finally, the dominance of the TLM approach diminishes with longer missing data segments, potentially related to its limitation in comparing templates solely in the projection space for extended intervals. Table \ref{tab:times_synth} provides average computation times for different algorithms, including the combined time of Steps 2 and 3 in $\mathsf{HaLI}$ for both interpolation alternatives. These findings highlight TLM's lower computational load and superior performance. 

\begin{table}[h]
\centering
\caption{Average computation time (in seconds) of different algorithms for the synthetic signal experiment.}
\small
\begin{tabular}{|c|c|c|c|c|c|}
	\hline
	& \multicolumn{4}{|c|}{$\mathbf{P_{ms} [\%]}$} \\
	\hline
	& \textbf{5} &\textbf{10} & \textbf{15} & \textbf{20}  \\
	\hline
	TLM & 0.027 & 0.0245 & 0.0235 & 0.0225 \\
	\hline
	LSE & 0.0116 & 0.01 & 0.001 & 0.001 \\
	\hline
	DMD & 0.025 & 0.0235 & 0.0216 & 0.0211  \\
	\hline
	EDMD & 0.09 & 0.13 & 0.09 & 0.09 \\
	\hline
	GPR & 1.690  & 1.531 & 1.455 & 1.412  \\
	\hline
	ARIMAF & 5.06 & 4.180 & 4.519 & 4.467 \\
	\hline
	ARIMAB & 4.707 & 4.477 & 4.286 & 4.010 \\
	\hline
	TBATS & 11.192 & 10.917 & 10.866 & 11.038 \\
	\hline
	DDTFA & 241.28 & 230.21 & 210.59 & 210.76 \\
	\hline
	LSW & 6.112 & 9.786 & 19.021 & 37.277 \\
	\hline
	\textsf{HaLI[BI](s)} (Steps 2+3) & 0.903 & 0.895 & 0.861 & 0.871 \\
	\hline
	\textsf{HaLI[BI](p)} (Steps 2+3) & 0.901 & 0.893 & 0.859 & 0.870 \\        
	\hline
\end{tabular}
\label{tab:times_synth}
\end{table}

\begin{table*}[h]
	\centering
	\caption{Median Normalized Mean Absolute Error (NMAE) and $p$-values of the signed-rank Wilcoxon tests for the initial imputation result and the \textsf{HaLI} using both interpolation schemes for all physiological signals. I: Best initial imputation. S: \textsf{HaLI[Best Imputation](s)}. P: \textsf{HaLI[Best Imputation](p)}. Statistically significant $p$-values are in boldface.}
	\label{tab:results_real}
	\small
	\begin{adjustbox}{angle=90}
		\begin{tabular}{|c|c|c|c|c||c|c|c|c|}
			\hline
			& \multicolumn{4}{c||}{\textbf{PPG}} & \multicolumn{4}{c|}{\textbf{ABP}}  \\
			\hline
			$\mathbf{P}_{ms}[\%]$ & \textbf{5} & \textbf{10} & \textbf{15} & \textbf{20} &  \textbf{5} & \textbf{10} & \textbf{15} & \textbf{20} \\
			\hline
			$\operatorname{NMAE}_I$ & $0.0572$ & $0.0829$ & $0.0958$ & $0.1087$ & $0.0440$ & $0.0619$ & $0.0659$ & $0.0793$ \\
			\hline
			$\operatorname{NMAE}_S$ & $0.0580$ & $0.0781$ & $0.0910$ & $0.1203$ & $0.0378$ & $0.0538$ & $0.0749$ & $0.0997$ \\
			\hline
			$\operatorname{NMAE}_P$ & $0.0561$ & $0.0752$ & $0.0832$ & $0.1053$ &  $0.0347$ & $0.0530$ & $0.0736$ & $0.0703$   \\
			\hline
			${p}_{I-S}$ & $\mathbf{0.016}$ & $\mathbf{0.0007}$ & $0.3533$ & $0.0780$ &  $\mathbf{0.0019}$ & $0.0524$ & $0.1546$ & $\mathbf{<0.0001}$ \\
			\hline
			${p}_{I-P}$ & $\mathbf{0.011}$ & $\mathbf{0.0004}$ & $0.0615$ & $0.1036$ &  $\mathbf{<0.0001}$ & $\mathbf{0.0018}$ & $\mathbf{0.0048}$ & $0.0697$ \\
			\hline
			${p}_{S-P}$ & $0.925$ & $0.3967$ & $0.0653$ & $\mathbf{0.0002}$ &  $\mathbf{<0.0001}$ & $\mathbf{0.0007}$ & $\mathbf{0.0003}$ & $\mathbf{<0.0001}$ \\
			\hline
			& \multicolumn{4}{c||}{\textbf{ACC}} & \multicolumn{4}{c|}{\textbf{AF}} \\
			\hline 
			$\operatorname{NMAE}_I$ & $ 0.0491$ & $ 0.0559$ & $ 0.0630 $ & $ 0.0702$ & $ 0.1344$ & $ 0.1361$ & $ 0.1742 $ & $ 0.2218$ \\
			\hline
			$\operatorname{NMAE}_S$ & $ 0.0468 $ & $ 0.0519 $ & $ 0.0633 $ & $  0.0766 $ & $ 0.1313 $ & $ 0.1197 $ & $ 0.1570 $ & $ 0.1806$ \\
			\hline
			$\operatorname{NMAE}_P$ & $ 0.0462 $ & $ 0.0504 $ & $ 0.0587 $ & $ 0.0664 $ & $ 0.1312 $ & $ 0.1211 $ & $ 0.1552 $ & $ 0.1788 $ \\
			\hline
			${p}_{I-S}$   & $ \mathbf{0.0225} $ & $ 0.0999 $ & $ 0.2699 $ & $ 0.8811 $ & $ 0.5228 $ & $ \mathbf{0.0003} $ & $ 0.0495 $ & $ \mathbf{0.0049} $ \\
			\hline
			${p}_{I-P}$   & $ \mathbf{0.0019} $ & $ \mathbf{0.0042} $ & $ \mathbf{0.0099} $ & $ \mathbf{0.0009} $ & $ 0.4925 $ & $ \mathbf{0.0003} $ & $ \mathbf{0.0010} $ & $ \mathbf{0.0003} $ \\
			\hline
			${p}_{S-P}$   & $ \mathbf{0.0029} $ & $ \mathbf{<0.0001} $ & $ \mathbf{<0.0001} $ & $ \mathbf{<0.0001}  $ & $ 0.5228 $ & $ 0.5228 $ & $ \mathbf{0.0016} $ & $ \mathbf{0.0086} $\\
			\hline
			& \multicolumn{4}{c||}{\textbf{NP}} & \multicolumn{4}{c|}{\textbf{THO}}\\
			\hline
			$\mathbf{P}_{ms}[\%]$ & \textbf{5} & \textbf{10} & \textbf{15} & \textbf{20} &  \textbf{5} & \textbf{10} & \textbf{15} & \textbf{20} \\
			\hline
			$\operatorname{NMAE}_I$ & $  0.1055 $ & $ 0.1233 $ & $ 0.1495 $ & $ 0.1568 $ & $  0.1065 $ & $ 0.1444 $ & $ 0.1620 $ & $ 0.2230 $ \\
			\hline
			$\operatorname{NMAE}_S$  & $  0.0987 $ & $ 0.1050 $ & $ 0.1404 $ & $ 0.1425 $ & $  0.0847 $ & $ 0.1211 $ & $ 0.1370 $ & $ 0.1859 $ \\
			\hline
			$\operatorname{NMAE}_P$  & $  0.0987 $ & $ 0.1024 $ & $ 0.1197 $ & $ 0.1320 $ & $  0.0840 $ & $ 0.1219 $ & $ 0.1260 $ & $ 0.1747 $ \\
			\hline
			${p}_{I-S}$  & $  \mathbf{0.0054} $ & $ \mathbf{0.0001} $ & $ \mathbf{0.0021} $ & $ \mathbf{0.0004} $ & $  0.141 $ & $ \mathbf{0.0001} $ & $ \mathbf{0.0022} $ & $ \mathbf{0.0002} $ \\
			\hline
			${p}_{I-P}$  & $ \mathbf{0.0049} $ & $ \mathbf{0.0001} $ & $ \mathbf{0.0001} $ & $ \mathbf{0.0001} $ & $  \mathbf{0.0141} $ & $ \mathbf{0.0001} $ & $ \mathbf{0.0001} $ & $ \mathbf{0.0001} $ \\
			\hline
			${p}_{S-P}$  & $  0.1500 $ & $ \mathbf{0.0027} $ & $ \mathbf{0.0001} $ & $ \mathbf{0.0004} $ & $ 0.0766 $ & $ \mathbf{0.0043} $ & $ \mathbf{0.0100} $ & $ \mathbf{0.0011} $ \\
			\hline
		\end{tabular}
	\end{adjustbox}
\end{table*}

\subsection{Missing data imputation in physiological signals.}
We analyzed real-world physiological signals from various sources, including the Taiwan Integrated Database for Intelligent Sleep (TIDIS)\footnote{https://tidis.org/en/}, the labeled raw accelerometry database\footnote{https://physionet.org/content/accelerometry-walk-climb-drive/1.0.0/}, and arterial blood pressure signals from the CHARIS database\footnote{https://physionet.org/content/charisdb/1.0.0/}. Our selection from the TIDIS database includes photoplethysmogram (PPG), airflow signal (AF), nasal pressure signal (NP), and thorax impedanciometry (THO) recordings. For accelerometry, we focused on the right ankle sensor's x-axis signal during walking. The CHARIS ABP signals exhibit a significant trend linked to mean arterial pressure (MAP), vital for the clinical assessment of imputed signals. In total, we evaluated $131$ distinct signals, distributed as $25$ PPG, $17$ AF, $25$ NP, $19$ THO, $32$ ACC, and $13$ ABP recordings. We selected segments free of saturation or sensor disconnections and with a clearly visible oscillatory pattern. For CHARIS, $39$ ABP segments from $3$ different selections were chosen. Each segment underwent three missing intervals, mirroring the approach from the previous section, with missing data rates ranging from $5\%$ to $20\%$. Notably, all signals comprise a solitary oscillatory component, excluding ABP where a nontrivial slow-varying oscillatory component tied to respiratory dynamics emerged. Thus, $K = 1$ was set for all signals except ABP, where $K = 2$ was chosen.

\begin{figure*}
\centering
\includegraphics[trim=0 10 0 0, width=\textwidth]{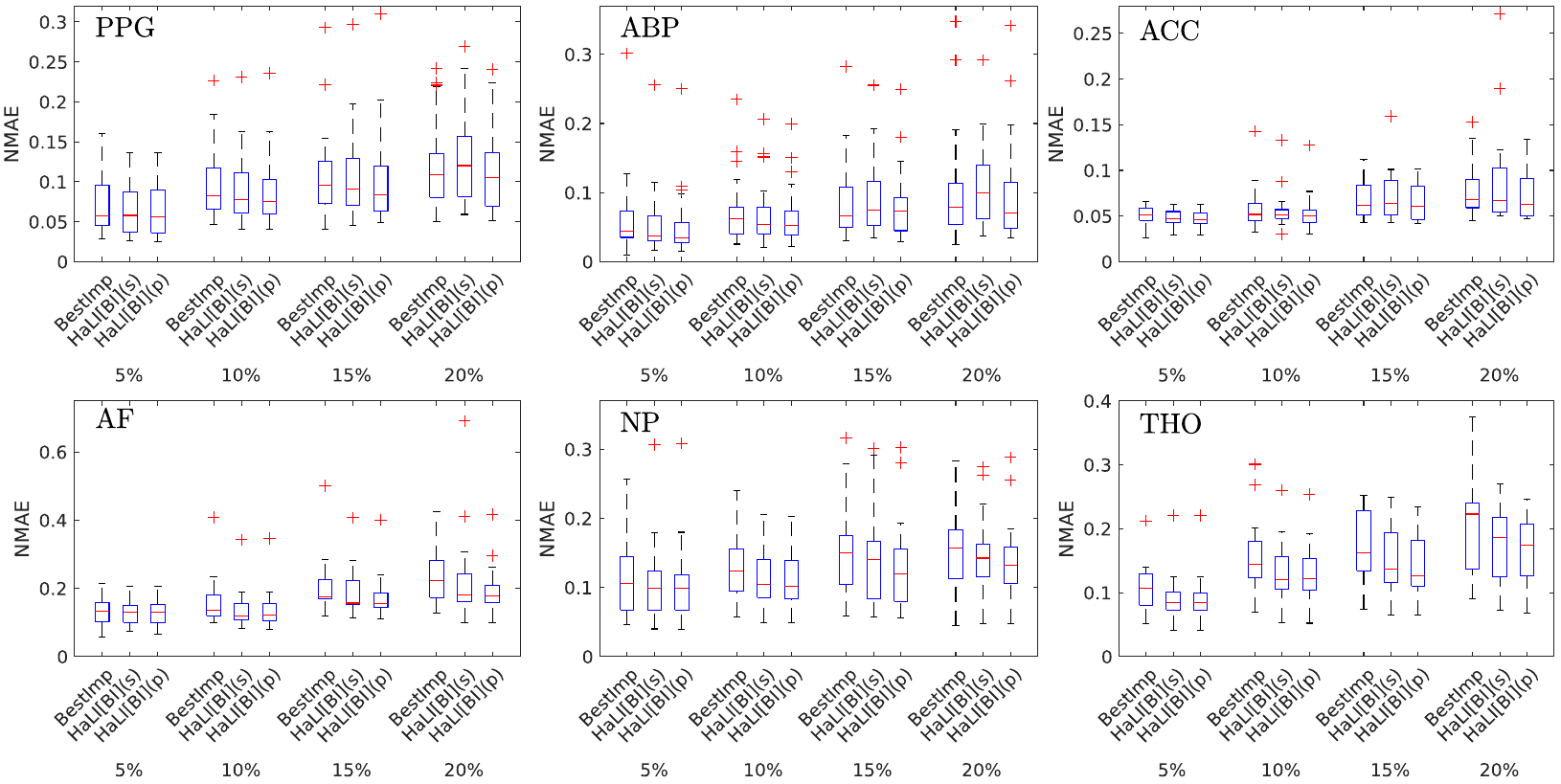}
\caption{Boxplots of the normalized mean absolute errors (NMAE) for each physiological signal considered at each missing data rate. BestImp: Best initial imputation; $\mathsf{HaLI[BI](s)}$: $\mathsf{HaLI}$ based on the best initial imputation and spline interpolation; $\mathsf{HaLI[BI](p)}$: $\mathsf{HaLI}$ based on the best initial imputation and shape-preserving cubic interpolation. Top: from left to right; result for PPG, ABP, and accelerometry. Bottom: from left to right; results for airflow, nasal pressure, and thorax impedanciometry.}
\label{fig:boxplot_real}
\end{figure*}

\begin{figure}
\centering
\includegraphics[trim=0 10 0 0, width=\columnwidth]{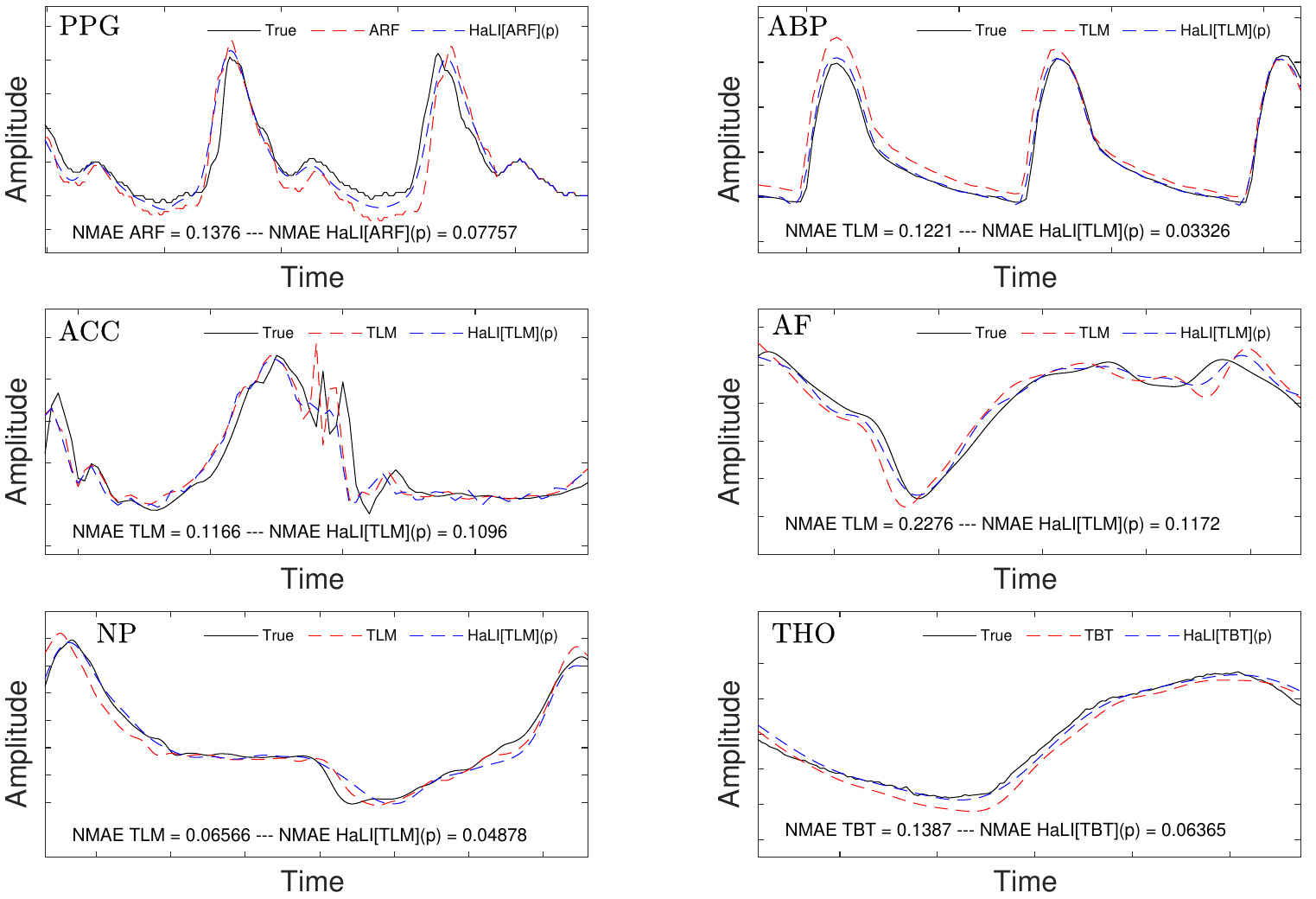}
\caption{Comparison between the initial missing data imputation before and after \textsf{HaLI} post-processing for each physiological signal. The original data is shown in black and the best initial imputation result and our improved imputation using \textsf{HaLI} are superimposed in red and blue, respectively. The $\operatorname{NMAE}$ with and without \textsf{HaLI} is shown at the bottom of each plot.}
\label{fig:curvesreal}
\end{figure}

\begin{figure}
\centering
\includegraphics[trim=0 10 0 0, width=\columnwidth]{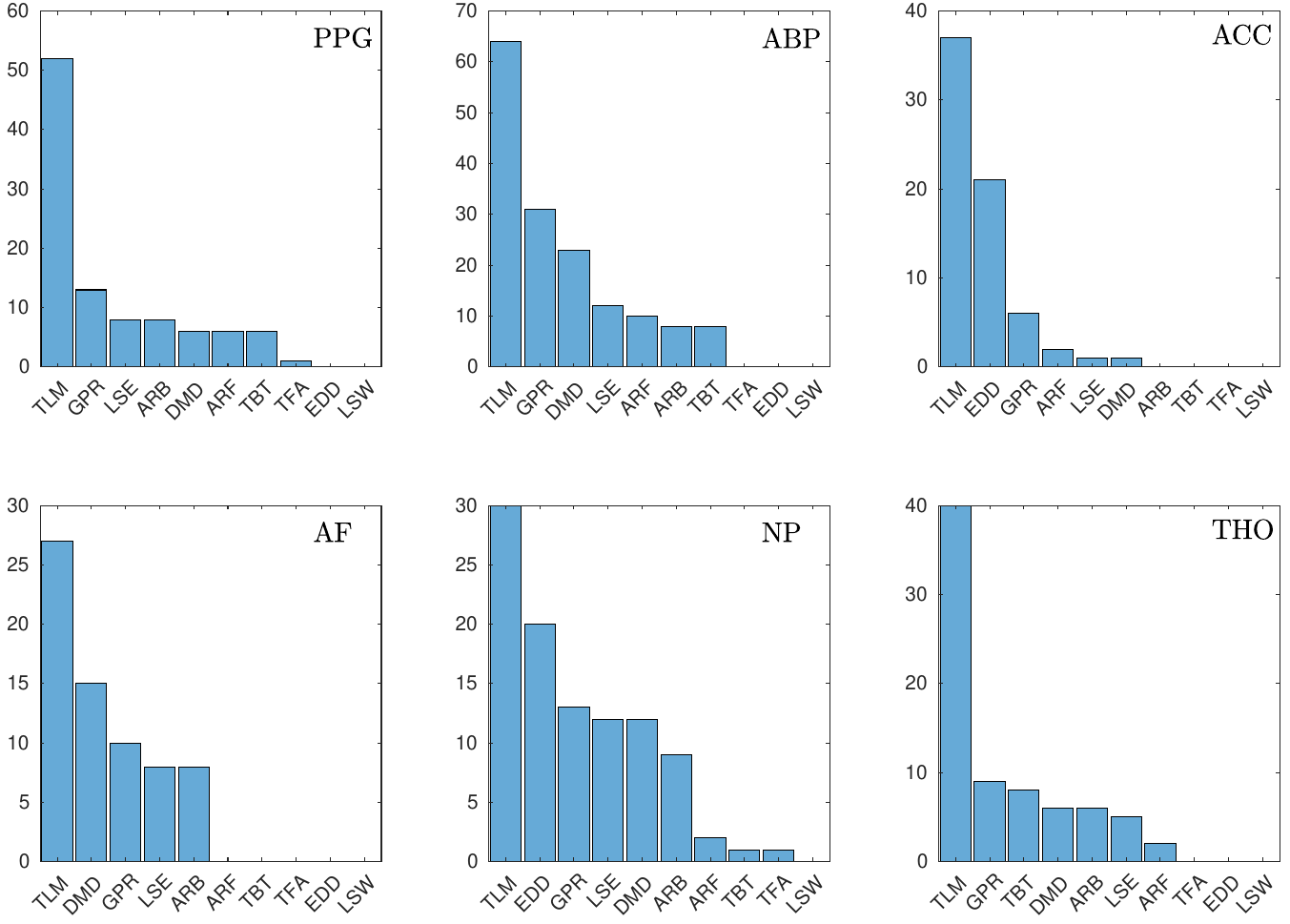}    
\caption{Imputation performance on real signals. Frequency histograms of the best performing initial imputation method on the different real signals considered. References: TLM: Takens' Lag Map. LSE: Least Square Estimation. DMD: Dynamic Mode Decomposition. EDD: Extended Dynamic Mode Decomposition. GPR: Gaussian Process Regression. ARF: ARIMA Regression with Forward Forecasting. ARB: ARIMA Regression with Backward Forecasting. TBT: Trigonometric Box-Cox, ARMA, and Seasonal Forecasting. TFA: Data-Driven Time-Frequency Analysis. LSW: Locally Stationary Wavelet Process.}
\label{fig:hist_real}
\end{figure}

The most effective combination for \textsf{HaLI} is determined based on error metrics, considering normalized mean absolute error ($\operatorname{NMAE}$) due to the varied magnitudes of physiological signals. The results are summarized in the boxplots in Fig. \ref{fig:boxplot_real}. Notably, the median $\operatorname{NMAE}$ value of \textsf{HaLI} is consistently lower compared to initial imputation in the majority of instances (as summarized in Table \ref{tab:results_real}). The Wilcoxon signed-rank hypothesis test confirms the statistical significance of this enhancement, as indicated by corresponding $p$-values in Table \ref{tab:results_real}. Across various $P_{ms}$ values and signals, \textsf{HaLI(p)} consistently improves the $\operatorname{NMAE}$ measure over initial imputation, except for specific cases such as PPG and ABP at $15\%$ and $20\%$, as well as AF signals at $5\%$ missing data rates. This is attributed to the significantly higher fundamental frequencies of PPG and ABP signals compared to respiratory signals (AF, NP, and THO), resulting in more cycles within the missing value interval(s). This underscores that initial imputation methods within this study face constraints when dealing with a high volume of consecutively missing signal cycles. Nevertheless, significant improvement is evident across other scenarios when applying the proposed approach. Fig. \ref{fig:curvesreal} offers a visual contrast between initial imputation and \textsf{HaLI}. We see that \textsf{HaLI} mitigates amplitude overshooting through amplitude interpolation and enhances synchronization between original and imputed cycles via phase interpolation. Subsequently, Fig. \ref{fig:hist_real} presents histograms for the top-performing initial imputation method across each signal, with TLM demonstrating superiority over other methods in the majority of cases.

\section{Discussion and conclusions}\label{sec:conclusions}

This study introduces a novel strategy for imputing missing data within non-stationary oscillatory time series. Our proposed method combines established imputation techniques, harmonic decomposition, and harmonic-level interpolation. We evaluate various missing data imputation algorithms and provide accessible Matlab code for wider use. Our experiments, involving synthetic and real-world signals, highlight the effectiveness of the \textsf{HaLI} approach, especially in harmonic-level interpolation. We analyze real signals with different sampling rates and fundamental frequencies, demonstrating the method's robustness across signal sample counts and cycle durations. In summary, \textsf{HaLI} shows great promise, with future work focused on exploring its clinical applications. Importantly, the harmonic decomposition and harmonic-level interpolation can be applied as post-processing steps to enhance any initial imputation method, offering versatility for integration into existing techniques.

Several technical considerations warrant discussion. First, real-world scenarios often involve mode-mixing setups, where the IFs of different IMT functions overlap. Our current algorithm has limitations in handling this issue, and we plan to explore new approaches for harmonic decomposition. The synchrosqueezed chirplet transform \cite{chen2023disentangling} shows promise in addressing this challenge. 
Next, the assumption of $B_{k,\ell}$ being a summable sequence indexed by $\ell$ mandates a non-spiky, oscillatory pattern. However, many real-world signals, like electrocardiograms or local field potentials with stimulation artifacts, exhibit spiky patterns. Adapting the proposed imputation algorithm to accommodate such scenarios is essential. 
Additionally, our current model characterizes oscillatory time series and their trends by describing their oscillatory behavior---a framework termed a phenomenological model. Enhanced knowledge of the background could warrant a more intricate statistical modeling of the IMT functions and trends. However, this tailored approach hinges on the application at hand and lies beyond the scope of this paper. 

This study narrows its focus to univariate time series. In practice, multivariate time series with missing values are common \cite{honaker2010missing, burger2018deriving}. While separate imputation is feasible, a pertinent question arises: Can missing values be collectively imputed? Examples of joint imputation can be found in \cite{weber2010imputation}. Within the statistical community, multiple imputation \cite{rubin2018multiple} is a well-established technique. When an oscillatory time series conforms to a statistical model akin to the framework in \cite{rubin2018multiple}, we conjecture that \textsf{HaLI} could be extended accordingly. If we viewed a time series as a function on $\mathbb{R}$, it is natural to ask if the proposed imputation idea could be generalized to spatial datasets, which is a function on $\mathbb{R}^2$ \cite{paiva2014imputation}, or even higher dimensional nonlinear spaces \cite{gilbert2018unsupervised}. These intriguing avenues will be the focus of our future investigations.

It is worth noting that various user-friendly software packages have been developed to facilitate missing time series imputation efficiently. Examples include R-based packages like 'imputeTS'\footnote{https://cran.r-project.org/web/packages/imputeTS/index.html} and Python-based packages like `impyute'\footnote{https://impyute.readthedocs.io/en/master/}, among others. These platforms offer a range of imputation methods, from simple ones like mean and last observation carried forward (LOCF) imputation to more advanced techniques based on non-stationary modeling, such as ARMA and structural time-series methods. Additionally, they often provide tools for data visualization and performance metric calculation. However, it is important to note that a readily available imputation toolbox in Matlab is currently unavailable. Therefore, this work contributes a publicly accessible Matlab implementation of the proposed imputation approach, serving as a practical module for Matlab users.

\bibliographystyle{IEEEtran}
\bibliography{imputation}

\end{document}